\documentclass[letterpaper,onecolumn]{quantumarticle}
\pdfoutput=1
\usepackage{amsmath,amsthm,amsfonts,amssymb}
\usepackage{graphicx}

\usepackage{color}
\usepackage{hyperref}
\newtheorem{theorem}{Theorem}
\newtheorem{corollary}{Corollary}
\newtheorem{lemma}{Lemma}

\newtheorem{assumption}{Assumption}
\usepackage{algorithm}

\DeclareMathAlphabet{\mathpzc}{OT1}{pzc}{m}{it}

\newcommand{\be}{\begin{equation}}
\newcommand{\ee}{\end{equation}}

\newcommand{\expec}{\mathbb{E}}

\newcommand{\Val}{{\rm Val}}

\newcommand{\vsig}{v_{\rm sig}}
\newcommand{\Tspike}{T_0}
\newcommand{\nbos}{{n_{bos}}}

\newcommand{\rhoone}{a^\dagger_\mu a_\nu}
\newcommand{\psirand}{\Psi_{\rm random}}
\newcommand{\dn}{D(N,\nbos)}
\newcommand{\psitarget}{\psi_{\rm target}}
\newcommand{\emax}{E_{max}}
\newcommand{\dem}{E'}
\newcommand{\ecut}{E_{cut}}
\newcommand{\Psisig}{\Psi_{\rm sig}}
\newcommand{\psiinput}{\Psi_{\rm input}}

\newcommand{\prob}{{\rm Pr}}
\newcommand{\Var}{{\rm Var}}
\newcommand{\corr}{{\rm corr}}
\newcommand{\epe}{E_{PE}}
\begin{document}

\title{Classical and Quantum Algorithms for Tensor Principal Component Analysis}

\author{Matthew B.~Hastings}

\affiliation{Station Q, Microsoft Research, Santa Barbara, CA 93106-6105, USA}
\affiliation{Microsoft Quantum and Microsoft Research, Redmond, WA 98052, USA}
\begin{abstract}
We present classical and quantum algorithms based on spectral methods for a problem in tensor principal component analysis.
The quantum algorithm achieves a {\it quartic} speedup while using exponentially smaller space than the fastest classical spectral algorithm, and a super-polynomial speedup over classical algorithms that use only polynomial space.
The classical algorithms that we present are related to, but slightly different from those presented recently in Ref.~\cite{wein2019kikuchi}.  In particular, we have an improved threshold for recovery and the algorithms we present work for both even and odd order tensors.  These results suggest that large-scale inference problems are a promising future application for quantum computers.
\end{abstract}
\maketitle

\section{Introduction}
Principal component analysis is a fundamental technique that finds applications in reducing the dimensionality of data and denoising.  While an optimal choice of principal components for a matrix can be computed efficiently using linear algebra,
the corresponding problem for tensors is much less well-understood.
Ref.~\cite{richard2014statistical} introduced 
a simple statistical model for tensor principal component analysis, termed 
the ``spiked tensor" problem, and this paper has lead to a large amount of follow-up research.  The model consists of (see below for more precise definitions) randomly choosing some unknown ``signal vector" $\vsig\in {\mathbb R}^N$; then, the $p$-th order tensor
\be
\Tspike=\lambda \vsig^{\otimes p} + G
\ee
is formed, where $G\in ({\mathbb R^N})^{\otimes p}$ is noise chosen from some random distribution and where $\lambda$ is some scalar representing a signal-to-noise ratio.  One task called recovery is to infer $\vsig$ (to some accuracy) given $\Tspike$.  A simpler task called detection is to distinguish the case $\lambda=0$ from $\lambda=\overline \lambda$ for some $\overline \lambda>0$, again just given $\Tspike$.

Ref.~\cite{richard2014statistical} presented a variety of algorithms for this problem.  
Following the normalization of Ref.
~\cite{wein2019kikuchi}, the entries of $G$ are chosen independently from a Gaussian distribution of zero mean and unit variance, with $|\vsig|=\sqrt{N}$.  
Then, information theoretically, it is possible to recover for $\lambda$ much larger than $N^{(1-p)/2}$~\cite{richard2014statistical,Lesieur_2017}.  However, no polynomial time algorithm is known that achieves this performance.
Rather, the two best known algorithms are spectral and sum-of-squares.
Spectral algorithms were first suggested in Ref.~\cite{richard2014statistical}.  There, a matrix is formed from $\Tspike$ (if $p$ is even, the matrix is $N^{p/2}$-by-$N^{p/2}$, with its entries given by entries of $\Tspike$) and the leading eigenvector of the matrix is used to determine $\vsig$.  For even $p$, this method works for $\lambda $ much larger than $N^{-p/4}$, and a variant of it is conjectured to perform similarly for odd $p$.  Methods based on the sum-of-squares also perform similarly to the spectral method.
The sum-of-squares method~\cite{hopkins2015tensor,Hopkins_2016} for this problem gives rise to a sequence of algorithms~\cite{tradeoff,tradeoff2}, in which one can recover at $\lambda$ smaller than $N^{-p/4}$ at the cost of runtime and space increasing exponentially in ${\rm polylog}(N) N^{-p/4}/\lambda$.
In Ref.~\cite{wein2019kikuchi}, a sequence of spectral algorithms with similar performance was shown.

In this paper, we present another spectral algorithm for the problem.  Our spectral algorithm for even $p$ is closely related to that of Ref.~\cite{wein2019kikuchi} which we became aware of while this paper was in preparation, and we use the normalization in that paper.  However, we make several changes. 
Our main technical results are the following.
First,
we prove an improved threshold for recovery for even $p$ using our algorithm; the improvement is by a constant factor and relies on a randomized method of recovering.  Second, we provide a different algorithm for odd $p$ with provable guarantees on while no guarantees were given in Ref.~\cite{wein2019kikuchi} for odd $p$.
For both even and odd $p$, we have
provable bounds on recovery for $\lambda$ of order $N^{-p/4}$ (without any polylogarithmic factors) and
we have a sequence of algorithms similar to that above for $\lambda$ small compared to $N^{-p/4}$.
Third, we give a quantum algorithm for our spectral method, achieving a quartic speedup and exponential reduction in space.
This quantum algorithm involves two main ideas.  The first uses phase estimation and amplitude amplification to obtain a quadratic speedup in computing the largest eigenvector.  The second idea uses a chosen input state to obtain a further quadratic speedup, giving the overall quartic speedup.

We emphasize that the quantum speedup is quartic compared to classical spectral algorithms presented here and in previous work.  We are not able to make an accurate comparison of the runtime to sum-of-squares methods.  In part, given that the runtime of all of these algorithms increases exponentially in $\lambda^{-1}$, a change in prefactors in some estimates for threshold can give rise to polynomial changes in runtime.  We expect that many of these estimates of thresholds are not tight (indeed, we expect that they are off by a polylogarithmic factor), and so either improved analytic methods or numerical simulations are needed to give an accurate comparison.

At a more heuristic level, we present a rather different motivation for our spectral algorithm compared to Ref.~\cite{wein2019kikuchi}.
Rather than being motivated by the so-called Kikuchi free energy, we instead are motivated by mean-field approximations to quantum many-body systems.  We consider a system of a some number $\nbos$ of qudits, each of dimension $N$, and use the tensor $\Tspike$ to construct a quantum Hamiltonian on these qudits. 
Increasing $\nbos$ gives rise to a similar sequence of algorithms as above, with increased runtime but improved performance: the required $\nbos$ increases polynomially in $\lambda^{-1}$ as ${\rm polylog}(N) (N^{-p/4}/\lambda)^{4/(p-2)}$, but the runtime increases exponentially.

 Restricting to the symmetric subspace, these $\nbos$ qudits can be thought of as a system of bosons.  In the case $p=4$, for example, our Hamiltonian has pairwise interaction terms for all pairs of qudits.  It is natural from the viewpoint of mean-field theory in physics then to expect that the leading eigenvector of the problem, for large $\nbos$, can be approximated by a product state.
While the bounds for arbitrary pairwise Hamiltonians would require rather large $\nbos$ for given $N$ in order for such a mean-field approximation to be accurate~\cite{Werner_1992,Kraus_2013,B2016}, we will be able to prove that for the statistical model above the mean-field approximation becomes accurate with high probablity at much smaller $\nbos$, depending upon the value of $\lambda$.
In this mean-field regime, the product state is an $\nbos$-fold tensor product of a single particle state, and this single particle state is given by $\vsig$ in an obvious way, regarding the vector $\vsig$ as a vector in the single-particle Hilbert space.
While we will not prove that this state is a good approximation to the leading eigenvector, it will be a good approximation to some state in an eigenspace with large eigenvalue.
Then, the single particle density matrix allows one to infer $\vsig$ (a similar matrix was used in Ref.~\cite{wein2019kikuchi} where it was termed a voting matrix).

Classically, implementing this spectral algorithm requires high-dimensional linear algebra, in particular finding the leading eigenvector of a matrix of dimension $\approx N^\nbos$.  This makes it a natural candidate for a quantum algorithm.  Since the Hamiltonian here is fairly simple, it can be simulated efficiently using standard techniques in the literature reviewed later.  This allows us to give a simple algorithm based on preparing a random initial state and then phase estimating in an attempt to project onto the leading eigenvector.  The probability of success in this projection is inverse in the dimension of the matrix, so this simple algorithm leads to no speedup over classical.  However, we show that it is possible to apply amplitude amplification to give a quadratic speedup over classical.  More surprisingly, we show that one can use the tensor $\Tspike$ to prepare an input state to the algorithm with improved overlap with the leading eigenvector, giving the quantum algorithm a quartic speedup over classical.
Here, when comparing to classical we are considering classical algorithms based on the power method or similar algorithms such as Lanczos; these algorithms require exponential space while the quantum algorithm uses only polynomial space.  We also consider classical algorithms based on ideas in Ref.~\cite{aaronson2017complexity} which use polynomial space but the quantum algorithm is super-polynomially faster than these algorithms.  We also present some minor improvements to the quantum algorithm which may be useful in practice.

\subsection{Definitions, Random Ensembles, and Notation}
Let us make some formal definitions.
A tensor $T$ of {\it order} $p$ and {\it dimension} $N$ is a multi-dimensional array.  The entries of the tensor are written $T_{\mu_1,\mu_2,\ldots,\mu_p}$ where $p \geq 1$ is an integer and each $\mu_a$ ranges from $1,\ldots,N$.
Generalizing previous work on this problem, we consider two possible cases, one in which entries of a tensor are chosen to be real numbers, and one in which they may be complex numbers, so that either $T\in ({\mathbb R^N})^{\otimes p}$ or
$T\in ({\mathbb C}^N)^{\otimes p}$; we explain later the reason for this generalization; a tensor with all entries real will be called a real tensor.
A symmetric tensor is one that is invariant under permutation of its indices.  The symmetrization of a tensor is equal to $1/p!$ times the sum of tensors given by permuting indices.

The spiked tensor model for given $N,p$ is defined as follows.
Let $\vsig$ be a vector in ${\mathbb R}^N$, normalized by $| \vsig| =\sqrt{N}$, chosen from some probability distribution; this is the ``signal vector".  Let $G$ be a real tensor of order $p$ with entries chosen from a Gaussian distribution with vanishing mean.
We let
$\Tspike=\lambda \vsig^{\otimes p} + G$ as above,
where $\vsig^{\otimes p}$ is defined to be the tensor with entries $$(\vsig^{\otimes p})_{\mu_1,\ldots,\mu_p}=\prod_{a=1}^p (\vsig)_{\mu_a}.$$
Here we use the notation that a subscript on a vector denotes an entry of that vector; we use a similar notation for matrices later.

Remark: some of the best sum-of-squares results are for a different distribution in which the entries of $\Tspike$ are chosen from a biased distribution on $\{-1,+1\}$, rather than for the Gaussian distribution.  We expect that using that distribution would not affect the results here too much, but we avoid treating that case also for simplicity.

Since the tensor $\vsig^{\otimes p}$ is symmetric, of course it is natural to replace $\Tspike$ by its symmetrization.  Indeed, no information can be lost by this replacement since given a tensor $\Tspike$ one can symmetrize the tensor, and then add back in Gaussian noise chosen to vanish under symmetrization to obtain a tensor drawn from the same distribution as $\Tspike$ was.
That is, the cases in which $G$ is symmetrized or not can be reduced to each other.

A generalization of this problem is the case in which $G$ is chosen to have complex entries, with each entry having
real and imaginary parts chosen from a Gaussian distribution with vanishing mean and variance $1/2$.  We refer to this as the complex ensemble, while we refer to the case where $G$ has real entries as the real ensemble; the choice of reducing the variance to $1/2$ is a convenient normalization for later.  It is clear that since $\vsig$ is real, the case of complex $G$ can be reduced to the real case (up to an overall rescaling for the different variance) simply by taking the real part of $\Tspike$, and similarly the real case can be reduced to the complex case (again up to an overall rescaling) by adding Gaussian distributed imaginary terms to the entries of $\Tspike$.
We will see that for odd $p$,
 at least for reasons of analyzing the algorithms, it is convenient not to symmetrize $\Tspike$ and to take complex $G$, while for even $p$ this is not necessary.  It may be possible to avoid doing this for odd $p$ (which may improve the detection and recovery threshold of the algorithm by constant factors) and we comment on this later.

We treat $p$ as fixed in the asymptotic notation, but consider the dependence on $N,\nbos$.
So, throughout this paper, when we refer to a polynomial in $N$, we mean a polynomial independent of the parameter $\nbos$.  
The polynomial may, however, depend on $p$, such as $N^p$.  

We make additionally the following assumptions:
\begin{assumption}
\label{assmp}
We assume that $\nbos=O(N^\theta)$ for some $p$-dependent constant $\theta>0$ chosen sufficiently small.
We will also assume that $\lambda$ is $\Omega(N^{-\theta'})$ for some $p$-dependent constant $\theta'>p/4$.

Finally, we assume that $\lambda=O(N^{-p/4})$.  Remark: there is of course no reason to consider $\lambda$ larger than this since simple spectral methods succeed if $\lambda$ is $\omega(N^{-p/4})$, but we state this assumption explicitly as it simplifies some of the big-O notation.
\end{assumption}
We will explicitly state this Assumption \ref{assmp} in all theorems where it is needed; the assumption will be implicit in the statement of the lemmas and will not be explicitly stated to avoid cluttering the statement of the results.

The first of these assumptions, that
$\nbos=O(N^\theta)$, is useful to simplify some of the statements of the results to avoid having to specify the allowed range of $\nbos$ in each case.  For example, we will say that a quantity such as $\nbos^p/N$ is $o(1)$, meaning that we must take $\theta<1/p$.  We do not specify the exact value of $\theta$ but it can be deduced from the proofs if desired.

The second of these assumptions,
that $\lambda$ is $\Omega(N^{-\theta'})$, also helps simplify some of the statements of the results.
Since we have assumed that $\nbos=O(N^\theta)$ and we will see that the required $\nbos$ increases polynomially with $\lambda^{-1}$, this assumed lower bound on $\lambda$ is not a further restriction on our results.

We write $\expec[\ldots]$ to denote expectation values and $\prob[\ldots]$ to denote a probability.  Usually these are expectation values or probabilities over choices of $G$, though in some cases we consider expectation values over other random variables.  We use $\Vert \ldots \Vert$ to denote the operator norm of an operator, i.e., the largest singular value of that operator.  We use $|\ldots|$ to denote either the $\ell_2$ norm of a tensor or the $\ell_2$ norm of a quantum state.  All logarithms are natural logarithms.

We use $\langle \ldots | \ldots \rangle$ to denote inner products and use bra-ket notation both for vectors and for quantum mechanical states.

We will need to compute expectation values over random choices of $G$ and also compute expectation values of certain operators in quantum states, such as $\langle \psi | O | \psi\rangle$ for some state $\psi$ and operator $O$.  We refer to the latter as a {\it quantum expectation value} of $O$ in state $\psi$ to distinguish it from an expectation value over random variables.

\subsection{Outline}
In section \ref{recovery}, we review some results on recovery and boosting from Ref.~\cite{wein2019kikuchi} and present a randomized recovery procedure that will help in improving the recovery threshold.
In section \ref{spectral}, we
give spectral algorithms for the spiked tensor problem for the case of both even and odd $p$.
In that section, we present algorithms in terms of eigenvalues and eigenvectors of a matrix (more precisely, vectors in some eigenspace and quantum expectation values of operators in those vectors)
that we call a {\it Hamiltonian}.  We leave the method to compute these eigenvalues and expectation values for 
later in section \ref{QCA}, where we give classical and quantum algorithms for this and give time bounds for those algorithms.
In section \ref{randspectra}, we give some results on the spectra of random tensors needed for the analysis of these algorithms.  A key idea here is reducing the case of a $p$-th order tensor for odd $p$ to the case of a $q$-th order tensor for even $q=2(p-1)$.  One interesting corollary of this technique, see corollary \ref{rmtcorr}, is that for odd $p$ and for the minimal value of $\nbos$ we are able remove a logarithmic factor in some of the bounds (a similar logarithmic factor has been removed also using  in ~\cite{hopkins2015tensor} using what they termed an ``unfolding" algorithm).
An appendix \ref{CTN} gives an introduction to some techniques used to evaluate expectation values of tensors networks whose tensors are chosen from a Gaussian distribution; these techniques are used earlier in the paper.
In section \ref{networksection}, we further discuss tensor networks and use this to consider limitations of certain algorithms and also to explain further some of the motivation for this algorithm.  In section \ref{discussionsection}, we discuss some extensions of the results.

The proof of detection is in theorem \ref{detecteven} for the even $p$ case and \ref{detectodd} for the odd $p$ case.
The proof of recovery is in theorem \ref{weakerconv} for the even $p$ case and theorem \ref{weakerconvodd} for the odd $p$ case. 
The runtime bound for the fastest quantum algorithm is in
theorem \ref{impruntimethm}.  This theorem gives a quartic improvement in the runtime compared to the fastest classical spectral algorithm; 
more precisely the log of the runtime with the quantum algorithm divided by the log of the runtime of the classical algorithm approaches $1/4$ as $N\rightarrow \infty$ at fixed $N^{-p/4}/\lambda$.

\section{Recovery}
\label{recovery}
In this section we discuss recovery and define randomized procedures for recovery that will be useful in boosting the threshold.
Following the notation of
Ref.~\cite{wein2019kikuchi}, define the {\it correlation} between two vectors by a normalized overlap
\be
\corr(x,y)=\frac{|\langle x|y\rangle|}{|x| \cdot |y|}.
\ee

The goal of an algorithm is to produce a vector $x$ with large $\corr(x,\vsig)$.  
Note that we take the absolute value in the correlation, ignoring the sign.  For even $p$, the sign of $\vsig$ is irrelevant, while for odd $p$ it is easy, given a guess of $\vsig$ up to sign, to try both choices of sign and see which is most likely.

Strong recovery means that
$\corr(x,\vsig)=1-o(1)$.
Proposition 2.6 of Ref.~\cite{wein2019kikuchi}, which is noted in that reference as being implicit in Ref.~\cite{richard2014statistical}, shows how to ``boost" a weaker correlation to strong recovery.  It is shown that 
given a vector $u$ one can apply a single iteration of the the tensor power algorithm to obtain a new vector $x$ such that,
with high probability,
\be
\corr(x,\vsig) \geq 1-c \lambda^{-1} \corr(u,\vsig)^{1-p} N^{(1-p)/2},
\ee
where $c$ is a constant depending on $p$.
So, for any $\lambda\omega(N^{(1-p)/2}$, if $\corr(u,\vsig)=\Omega(1)$, we have $\corr(x,\vsig)=1-o(1)$.

Thus, it suffices to construct which outputs some vector $u$ which has, with high probability, $\corr(u,\vsig)=\Omega(1)$.
This is termed {\it weak recovery}; indeed, one can be satisfied with even weaker assumptions depending on the value of $\lambda$; for $\lambda$ close to $N^{-p/4}$, one may have polynomially small $\corr(u,\vsig)$.

The spectral algorithms that we construct later will output a matrix that we will write $\rho_1$.  This matrix will be
a positive semi-definite Hermitian matrix with trace $1$.  In physics terms, such a matrix is termed a {\it density matrix}.
For sufficiently large $\lambda$, the leading eigenvector of the matrix will have a large overlap with $\vsig$.  However, for smaller $\lambda$, we will still have some lower bound that, with high probability,
$\langle \vsig | \rho_1| \vsig \rangle=\Omega(1) |\vsig|^2$. 
The randomized algorithm \ref{randrec} below shows how to use this to obtain weak recovery.
This randomized algorithm allows us to improve the threshold over the recovery method~\cite{wein2019kikuchi} of simply using the leading eigenvector $\rho_1$ since it works even in some cases where the leading eigenvector has small correlation with $\vsig$.

Putting these results together we find that
\begin{corollary}
Given an algorithm that, with high probability, outputs a density matrix $\rho$ with
$$\frac{\langle \vsig | \rho_1 |\vsig \rangle}{N}=\Omega(1),$$
then in polynomial time, with high probability, strong recovery is possible.

We will simply say that such an algorithm {\it achieves recovery}.
\end{corollary}

We present the algorithm for the case that the matrix $\rho_1$ may be complex; if the matrix is real, one can instead sample from a real Gaussian distribution and the proof of lemma \ref{reclemma} goes through will slightly different constants.

\begin{algorithm}
\caption{Input: density matrix $\rho$.  Output, some vector $w$ obeying bounds described above}
\begin{itemize}
\item[1.] Randomly sample a vector $u$ with entries chosen from a correlated complex Gaussian distribution with zero mean and with covariance
$\expec[\overline u_i u_j]=(\rho)_{ij},$ with $\expec[u_i u_j]=\expec[\overline u_i \overline u_j]=0$.

\item[2.] Let $w=u/|u|$.
\end{itemize}
\label{randrec}
\end{algorithm}

We have
\begin{lemma}
\label{reclemma}
For algorithm \ref{randrec},
with probability at least $1/2$,
\be
|\langle w | \vsig \rangle| \geq c'\sqrt{\langle \vsig | \rho| \vsig\rangle},
\ee
for some scalar $c'>0$.

\begin{proof}
We have $\expec[|u|^2]={\rm tr}(\rho)=1$.
Hence, with probability at least $3/4$ we have that $|u|^2\leq 4/3$.
We have $\expec[|\langle u | \vsig \rangle|^2]=\langle \vsig | \rho \vsig\rangle$ and since 
$\langle u | \vsig \rangle$ is a Gaussian random variable with mean $0$, with probability at least $3/4$ its absolute value is
at least some positive constant $c''$ (the exact constant can be deduced from the error function) times its standard deviation.
Hence, the lemma follows for $c'=(3/4) c''$.
\end{proof}
\end{lemma}

\section{Spectral Algorithm}
\label{spectral}
We now give the
spectral algorithms for the spiked tensor problem for the case of both even and odd $p$.
In subsection \ref{Hamdef}, we define a Hamiltonian $H(T)$ given an arbitrary tensor $T$.
Then, in subsection \ref{spectralres}, we present the spectral algorithm in terms of $H(\Tspike)$.

For even $p$, the Hamiltonian that we present is very similar to the matrix $Y$ given in Ref.~\cite{wein2019kikuchi} but it has some minor differences.  In our language (explained below), this matrix $Y$ is obtained by projecting our Hamiltonian of Eq.~(\ref{firstquantized}) into the subspace of the ``symmetric subspace" spanned by $|\mu_1\rangle \otimes |\mu_2\rangle \otimes \ldots \otimes |\mu_{\nbos}\rangle$ with $\mu_1,\ldots,\mu_{\nbos}$ all distinct.  The relative reduction in the size of the matrix is only $O(1/N)$ in the limit of large $N$.

Also, in our method, we have an $O(N)$ rotational symmetry of the basis which is very useful in analysis, for example showing that the eigenvalues of $H(\lambda \vsig^{\otimes p})$ are independent of choice of $\vsig$.  For the matrix $Y$ of ~\cite{wein2019kikuchi}, this is not obvious to us and we do not fully understand the claimed behavior of the largest eigenvalue in that case.
We will use a different notation, using creation and annihilation operators, which will help make this rotational symmetry more explicit.

For odd $p$, the Hamiltonian that we use is unrelated to that of Ref.~\cite{wein2019kikuchi}.

\subsection{Hamiltonian Definition}
\label{Hamdef}
For even $p$, given a tensor $T$ we define an linear operator $H(T)$ that we call a {\it Hamiltonian} as follows.
This Hamiltonian is a linear operator on a vector space $({\mathbb R}^{N})^{\otimes \nbos}$ or $({\mathbb C}^{N})^{\otimes \nbos}$, for integer $\nbos\geq 1$ chosen below.  We write basis elements of this space as $|\mu_1\rangle \otimes |\mu_2\rangle \otimes \ldots \otimes |\mu_\nbos\rangle$, and we call this space the {\it full Hilbert space}.
We define
\be
\label{firstquantized}
H(T)  =\frac{1}{2}\sum_{i_1,\ldots,i_{p/2}}  \Bigl( \sum_{\mu_1,\ldots,\mu_p} T_{\mu_1,\mu_2,\ldots,\mu_p} 
|\mu_1\rangle_{i_1}\langle \mu_{1+p/2}| \otimes 
|\mu_2\rangle_{i_2}\langle \mu_{2+p/2}| \otimes  \ldots \otimes
|\mu_{p/2}\rangle_{i_{p/2}}\langle \mu_{p}| + {\rm h.c.} \Bigr),
\ee
where the sum is over distinct $i_1,i_2,\ldots,i_{p/2}$ so that there are $(p/2!) {\nbos \choose p/2}$ terms in the sum and
where ${\rm h.c.}$ means adding the Hermitian conjugate of the given terms, so that $H(T)$ is Hermitian and where
$|\mu\rangle_i\langle\nu|$ denotes the operator $|\mu\rangle\langle \nu|$ on qudit $i$.
We require that $\nbos\geq p/2$ or else $H(T)$ is trivially zero.

Note of course that if $T$ is real and symmetric, then the term
$ \sum_{\mu_1,\ldots,\mu_p} T_{\mu_1,\mu_2,\ldots,\mu_p} 
|\mu_1\rangle_{i_1}\langle \mu_{1+p/2}| \otimes 
|\mu_2\rangle_{i_2}\langle \mu_{2+p/2}| \otimes  \ldots \otimes
|\mu_{p/2}\rangle_{i_{p/2}}\langle \mu_{p}| $ is already Hermitian.
$H(T)$ can be regarded as a Hamiltonian acting on a space of $\nbos$ qudits, each of dimension $N$,
and with interaction between sets of $p/2$ particles at a time.

Even if $T$ is not symmetrized, $H(T)$ is unchanged if one applies an arbitrary permutation to the first $p/2$ indices of $T$ and applies the same permutation to the last $p/2$ indices of $T$.

We may restrict to the symmetric subspace of this Hilbert space.
We write
$D(N,\nbos)$ to indicate the dimension of this subspace.  For $N\gg \nbos$, we can approximate $D(N,\nbos) \approx N^\nbos/\nbos!$.

Within the symmetric subspace, we can write this Hamiltonian in a so-called ``second-quantized" form:
\be
\label{sqeven}
H(T)=\frac{1}{2} \Bigl(\sum_{\mu_1,\ldots,\mu_p} T_{\mu_1,\mu_2,\ldots,\mu_p} 
\Bigl(\prod_{i=1}^{p/2} a^{\dagger}_{\mu_i} \Bigr)
\Bigl(\prod_{i=p/2+1}^p  a_{\mu_i}\Bigr)+{\rm h.c.}\Bigr).
\ee
This replacement by a second-quantized Hamiltonian is simply a convenient notation.
The operators $a^\dagger_\mu,a_\mu$ are bosonic creation and annihilation operators, obeying canonical commutation relations
$[a_\mu,a^\dagger_\nu]=\delta_{\mu,\nu}$.  See appendix \ref{ccr} for a brief review of this formalism.
We restrict to the subspace with a total of $\nbos$ bosons, i.e.,
we define the number operator $n$ by
\be
n\equiv \sum_\mu a^\dagger_\mu a_\mu,
\ee
and restrict to $n=\nbos.$
An orthonormal basis of states for this symmetric subspace is given by all states equal to
some normalization constant multiplying $a^\dagger_{\mu_1} a^\dagger_{\mu_2} \ldots a^{\dagger}_{\mu_\nbos} | 0\rangle$, where $|0\rangle$ is the {\it vacuum} state (i.e., the state annihilated by $a_\mu$ for all $\mu$), and where $\mu_1\leq \mu_2 \leq \ldots \leq \mu_\nbos$ is some sequence.

The second quantized Hamiltonian for the symmetric subspace is unchanged under arbitrary permutation of the first $p/2$ indices of $T$ and arbitrary (not necessarily the same) permutation of the last $p/2$ indices of $T$.

For odd $p$, we define the Hamiltonian $H(T)$
as follows.  Given a tensor $T$ of odd order $p$, define a new tensor $\tilde T$ of even order $q=2(p-1)$ with components
\be
\label{tildeTdef}
\tilde T_{\mu_1,\ldots,\mu_{(p-1)/2},\nu_1,\ldots,\nu_{(p-1)/2},\mu_{(p-1)/2+1},\ldots,\mu_{p-1},\nu_{(p-1)/2},\ldots,\nu_{p-1}}=\sum_\sigma T_{\mu_1,\ldots,\mu_{p-1},\sigma} T_{\nu_1,\ldots,\nu_{p-1},\sigma}.
\ee
Then define $H(T)=H(\tilde T)$, using the definition (\ref{firstquantized}) for $H(\tilde T)$.
Note the order of indices on the left-hand side of Eq.~(\ref{tildeTdef}).
Using the second-quantized notation, this gives for odd $p$:
\begin{eqnarray}
\label{sqodd}
&&H(T) \\ \nonumber &=&\frac{1}{2} \Bigl(\sum_{\mu_1,\ldots,\mu_{p-1}} \sum_{\nu_1,\ldots,\nu_{p-1}}  \sum_{\sigma}
T_{\mu_1,\mu_2,\ldots,\mu_{p-1},\sigma} T_{\nu_1,\nu_2,\ldots,\nu_{p-1},\sigma}
\Bigl(\prod_{i=1}^{(p-1)/2} a^{\dagger}_{\mu_i} 
a^\dagger_{\nu_i} \Bigr)
\Bigl(\prod_{i=(p-1)/2+1}^{p-1}  a_{\mu_i}a_{\nu_i} \Bigr)+{\rm h.c.}\Bigr),
\end{eqnarray}
Now we require that $\nbos\geq p-1$ as otherwise $H(T)$ is trivially zero.
For this Hamiltonian, it is convenient to take $G$ from the complex ensemble because, as we explain more below, it makes $\expec[H(G)]$ equal to zero, as well as canceling out certain terms in higher order moments, making the proof of the spectral properties of $H(G)$ simpler.  We discus later to what extent we can avoid using the complex ensemble.

\subsection{Spectral Algorithms}
\label{spectralres}
The spectral algorithm for detection and recovery is algorithm \ref{specalg}.
In this subsection we prove correctness of this algorithm, using statistical properties of $H(G)$ proven later.

This algorithm uses quantities $E_0$ and $\emax$ defined later; roughly $\emax$ is an upper bound on the eigenvalues of $H(G)$ and $E_0$ is the largest eigenvalue of $H(\lambda v^{\otimes p})$.  The algorithm can then achieve detection by verifying that the largest eigenvalue is significantly larger than $\emax$, which occurs when $E_0$ is large enough.
Indeed, we will see that it suffices to have $E_0=(1+c) \emax$ for any {\it fixed} $c>0$ (some results can be extended to the case that $c$ decays slowly with $N$ but we omit this for brevity).  This fixes the scaling of $\nbos$ with $\lambda$ so that we need (up to polylogarithmic factors) $\nbos \gtrsim (N^{-p/4}/\lambda)^{4/(p-2)}$.

One interesting feature of the algorithm is that in step 3, we compute the density matrix of the leading eigenvector or of any vector in the eigenspace of eigenvalue $\geq \ecut$, for  $\ecut=(E_0+\emax)/2$ defined in the algorithm.  This might seem surprising, given that the leading eigenvector is computed in step 1; one might wonder why some other vector should be taken.  We describe the algorithm in this way since, in later classical and quantum algorithms that we give to compute the
spectral properties of the matrix, we might not extract the leading eigenvector but instead extract only some vector in this eigenspace due to use of the power method in a classical algorithm or due to approximations in phase estimation in a quantum algorithm. 
Thus,
much of our analysis is given to showing that {\it some} eigenvalue larger than $\ecut$ exists by lower bounding the leading eigenvalue $\lambda_1$, but given that some such eigenvalue exists, we do not worry too much about exactly what mixture of eigenvectors in the given eigenspace we compute.

\begin{algorithm}
\caption{Spectral algorithm.  This algorithm takes a tensor $\Tspike$ as input and also a scalar $\overline \lambda$ and an integer $\nbos$.  The output is a decision about whether $\lambda=\overline \lambda$ or $\lambda=0$, and, if the algorithm reports that $\lambda=\overline \lambda$, it also returns an approximation of $\vsig$ (up to an overall sign).
The quantity $\nbos$ is chosen depending upon the value of $\overline \lambda$; smaller values of $\lambda$ require larger values of $\nbos$ in order for $E_0$ to be sufficiently larger than $\emax$ for the algorithm
to be accurate.  See theorems \ref{detecteven},\ref{detectodd},\ref{weakerconv},\ref{weakerconvodd}.  For $E_0\geq (1+c) \emax$ for any $c>0$, the algorithm achieves recovery.}
\begin{itemize}
\item[1.] Compute the eigenvector of $H(\Tspike)$ and the leading eigenvalue, denoted $\lambda_1$.

\item[2.] (Detection) If $$\lambda_1>\ecut\equiv (E_0+\emax)/2,$$ 
where $E_0=\overline \lambda (p/2)! {\nbos \choose p/2} N^{p/2}$ for even $p$, and $E_0=\overline \lambda^2 (p-1)! {\nbos \choose p-1} N^p$ for odd $p$, and
where $\emax$ is defined in theorem \ref{eigbthm}, then report that $\lambda=\overline \lambda$.  Otherwise report $\lambda=0$.

\item[3.] (Recovery) Compute the single particle density matrix (defined below) of the leading eigenvector or of any vector in the eigenspace of eigenvalue $\geq \ecut$.  Apply algorithm \ref{randrec} to recover an approximation to $\vsig$.
\end{itemize}
\label{specalg}
\end{algorithm}

In section \ref{randspectra}, we consider the largest eigenvalue $\lambda_1$ of $H(G)$ and show the following theorem which summarizes the results of lemmas \ref{eigbound},\ref{eigboundodd}.  Roughly, up to prefactors, the result for odd $p$ is given by considering the result for a $q$-th order tensor for even $q=2(p-1)$ and then multiplying the largest eigenvalue by a factor of $\sqrt{N}$.
\begin{theorem}
\label{eigbthm}
Let $\lambda_1$ be the largest eigenvalue of $G$.
For even $p$ let $$\emax=\sqrt{2J \log(N)} \nbos^{p/4+1/2} N^{p/4},$$
and for odd $p$ let
$$\emax=2\sqrt{J \log(N)} \nbos^{p/2} N^{p/2},$$
where 
$J$ is a scalar depends that implicitly on $p,\nbos,N$ and
tends to some function depending only on $p$ for large $\nbos,N$.  More precisely, for even $p$, $J$ 
 is equal to $(p/2)! {\nbos \choose p/2!}/\nbos^{p/2}+o(1)$ for the real ensemble and is twice that for the complex ensemble, and for odd $p$, $J$ is equal to that for the even case for $2(p-1)$.

Then,
for any $x$, assuming Assumption \ref{assmp},
\be
\prob[\lambda_1\geq x] \leq \exp\Bigl(-\frac{x-\emax}{\xi}\Bigr),
\ee
with for even $p$
\be
\xi=\frac{\sqrt{J} \nbos^{p/4-1/2} N^{p/4}}{\sqrt{2\log(N)}}
\ee
and for odd $p$
\be
\xi=\frac{\sqrt{J} \nbos^{p/2-1} N^{p/2}}{\sqrt{\log(N)}}.
\ee

So, for any $\dem$ which is $\omega(\xi)$, with high probability $\lambda_1\leq \emax+\dem$.
\end{theorem}

Now consider the eigenvectors and eigenvalues of $H(\Tspike)$.  
For any symmetric tensor $T$ of order $\nbos$, let $|T\rangle$ be
the vector on $\nbos$ qudits (each of dimension $N$) with amplitudes given by 
the entries of the tensor in the obvious way:
$$|T\rangle=\sum_{\mu_1,\ldots,\mu_{\nbos}} T_{\mu_1,\ldots,\mu_{\nbos}}
|\mu_1\rangle \otimes \ldots \otimes |\mu_{\nbos}\rangle.$$
This vector is only normalized if $|T|=1$.
So, $\Psisig\equiv N^{-\nbos/2} |\vsig^{\otimes \nbos}\rangle$ is a normalized vector.

We have the following simple property:
\begin{lemma}
\label{collect}
Let $\lambda_1$ be the largest eigenvalues of $H(\Tspike)$.  
Then, $\lambda_1 \geq E \equiv \langle \Psisig | H(T) |\Psisig \rangle$.
\begin{proof}
Immediate from the variational principle.
\end{proof}
\end{lemma}

\subsubsection{Even $p$ Case}
\label{evenpc}
We now show correctness of the algorithm.
All results in this subsubsection refer to even $p$ even if we do not state it explicitly.
First we estimate $E$ of lemma \ref{collect} to show detection.  Then, we show recovery.

We have
\begin{eqnarray}
E& =& \langle\Psisig | H(\lambda \vsig^{\otimes p}) | \Psisig \rangle
+\langle\Psisig | H(G) | \Psisig \rangle
 \\
&=&E_0+
\langle\Psisig | H(G) | \Psisig \rangle,
\end{eqnarray}
where
\be
E_0=\lambda (p/2)! {\nbos \choose p/2} N^{p/2}.
\ee
To evaluate
$\langle\Psisig | H(G) | \Psisig \rangle$
it is convenient to exploit a rotational invariance of this problem.  We can apply a rotation using any matrix $O$ in the orthogonal group $O(N)$,
rotating the creation and annihilation operators by making the replacement $$a^\dagger_{\mu} \rightarrow \sum_{\nu} O_{\mu \nu} a^\dagger_\nu$$ and
$$a_{\mu} \rightarrow \sum_{\nu} O_{\mu \nu} a_\nu.$$
This rotation preserves the canonical commutation relations and is equivalent to rotating the basis states on each qudit by $O$.
To preserve the Hamiltonian $H(\Tspike)$, we rotate each leg of the tensor $\Tspike$:
$$(\Tspike)_{\mu_1,\ldots,\mu_p} \rightarrow \sum_{\nu_1,\ldots,\nu_p} (\Tspike)_{\nu_1,\ldots,\nu_p}
\prod_i O_{\mu_i,\nu_i}.$$
This rotation preserves the Gaussian measure on $G$ but changes $\vsig$.
So, we can rotate so that $\vsig$ is some fixed vector, say $(\sqrt{N},0,0,\ldots)$ so $(\vsig)_1=\sqrt{N}$.  Then
$\langle\Psisig | H(G) | \Psisig \rangle$ is equal to $ (p/2)! {\nbos \choose p/2}$ multiplied by a single entry of $G$, i.e., the entries with all indices equal to $1$, which is some quantity chosen from a Gaussian distribution with zero mean and unit variance.
So, with high probability, $E=\lambda (p/2)! {\nbos \choose p/2} N^{p/2}+\nbos^{p/2} O(N^\eta)$ for any $\eta>0$.

Hence,
\begin{theorem}
\label{detecteven}
If $\lambda (p/2)! {\nbos \choose p/2} N^{p/2} - \emax=\omega\Bigl(\frac{\sqrt{J} \nbos^{p/4-1/2} N^{p/4}}{\sqrt{2\log(N)}}\Bigr)$, and if Assumption \ref{assmp} holds,
then with high probability algorithm \ref{specalg} correctly determines whether $\lambda=0$ or $\lambda=\overline \lambda$.
\begin{proof}
This follows from the estimate of $E$ which lowers bounds the largest eigenvalue when $\lambda=\overline \lambda$ and from theorem \ref{eigbthm} which upper bounds $\Vert H(G) \Vert$.
\end{proof}
\end{theorem}

Given any state, define the {\it single particle density matrix} $\rho_1$ in the basis for the full Hilbert space used in Eq.~(\ref{firstquantized}) to be the reduced density matrix on any of the qudits.  Equivalently, this single particle density matrix can be expressed in terms of creation and annihilation operators by
\be
(\rho_1)_{\mu \nu}=\frac{1}{\nbos} \rhoone.
\ee
Note that $\rho_1$ is positive semi-definite, Hermitian, and trace $1$.

We have
\begin{theorem}
\label{weakerconv}
Let Assumption \ref{assmp} hold.
Given any vector $\Psi$ such that $\langle \Psi | H(\Tspike) |\Psi\rangle \geq (1+c') \emax$ for any scalar $c'>0$,
then with high probability
the corresponding single particle density matrix
$\rho_1$ obeys
\be
\label{obeys}
\frac{\langle \vsig| \rho_1 |\vsig\rangle}{N} \geq (c'-o(1))\frac{\emax}{E_0}.
\ee

In particular, for $\langle \Psi | H(\Tspike) |\Psi\rangle \geq \ecut$ with $E_0\geq (1+c) \emax$ we have
$c'\geq c/2$
so with high probability
\be
\label{obeys2}
\frac{\langle \vsig| \rho_1 |\vsig\rangle}{N} \geq \frac{1}{2} \frac{c}{1+c}.
\ee

Hence, algorithm \ref{specalg} achieves recovery.

\begin{proof}
We have
$\langle \Psi | H(\lambda \vsig^{\otimes p} |\Psi\rangle+ \langle \Psi| H(G) |\Psi\rangle \geq (1+c) \emax$.
By theorem \ref{eigbthm}, with high probability  $\langle \Psi| H(G) |\Psi\rangle\leq (1+o(1)) \emax$.
Hence, with high probability $\langle \Psi | H(\lambda \vsig^{\otimes p}) |\Psi\rangle\geq (c-o(1)) \emax$.

Rotate so that $\vsig=(\sqrt{N},0,0,\ldots)$.
Then, the Hamiltonian $H(\lambda \vsig^{\otimes p})$ is diagonal in the computational basis for the full Hilbert space.
Let $P(m)$ be the probability, for state $\Psi$, that exactly $m$ out of $\nbos$ qudits are in state $|1\rangle$.  Then,
$\langle \Psi | H(\lambda \vsig^{\otimes p})| \Psi \rangle=\sum_m \lambda (p/2)! {m \choose p/2} N^{p/2} P(m)
\leq \sum_m m (E_0/\nbos) P(m)=E_0
\langle \vsig| \rho_1 | \vsig\rangle/N$.
\end{proof}
\end{theorem}
Remark: more generally, the same result holds for any mixed state: 
given any mixed state $\sigma$ such that ${\rm tr}(\sigma H(\Tspike))  \geq (1+c) \emax$ for any scalar $c>0$,
then with high probability
the corresponding single particle density matrix
$\rho_1$ obeys Eq.~(\ref{obeys}).

\subsubsection{Odd $p$ Case}
\label{oddpc}
We now show correctness of the algorithm for odd $p$.  All results in this subsubsection refer to odd $p$ even if we do not state it explicitly.

Let us first estimate $E$ to show detection.
We have
$E=\langle\Psisig | H(\Tspike) | \Psisig \rangle$,
but now $H(\Tspike)$ is a quadratic function of $\Tspike$ so there are some cross-terms.
We have
\be
\label{oddpexpect}
\langle\Psisig | H(\vsig^{\otimes p}) | \Psisig \rangle=\lambda^2 (p-1)! {\nbos \choose p-1} N^{p}\equiv E_0.
\ee

The cross-term is
$$\lambda \langle \Psisig |  \sum_{\mu_1,\ldots,\mu_{p-1}} \sum_{\nu_1,\ldots,\nu_{p-1}}  \sum_{\sigma}
(\vsig^\otimes p)_{\mu_1,\mu_2,\ldots,\mu_{p-1},\sigma} G_{\nu_1,\nu_2,\ldots,\nu_{p-1},\sigma}
\Bigl(\prod_{i=1}^{(p-1)/2} a^{\dagger}_{\mu_i} 
a^\dagger_{\nu_i} \Bigr)
\Bigl(\prod_{i=(p-1)/2+1}^{p-1}  a_{\mu_i}a_{\nu_i} \Bigr)+{\rm h.c.} |\Psisig\rangle.$$
Exploiting the same rotational invariance as in the even $p$ case, this is equal to 
$2 \lambda (p-1)! {\nbos \choose p-1}N^{p/2}$ multiplied by the real part of a single entry of $G$, i.e., the entry with all indices equal to $1$.
So, with high probability, this cross-term is bounded by $\lambda \nbos^{p-1} O(N^{p/2+\eta})$ for any $\eta>0$.

The term quadratic in $G$ is 
$$\frac{1}{2} \langle \Psisig |\sum_{\mu_1,\ldots,\mu_{p-1}} \sum_{\nu_1,\ldots,\nu_{p-1}}  \sum_{\sigma}
G_{\mu_1,\mu_2,\ldots,\mu_{p-1},\sigma} G_{\nu_1,\nu_2,\ldots,\nu_{p-1},\sigma}
\Bigl(\prod_{i=1}^{(p-1)/2} a^{\dagger}_{\mu_i} 
a^\dagger_{\nu_i} \Bigr)
\Bigl(\prod_{i=(p-1)/2+1}^{p-1}  a_{\mu_i}a_{\nu_i} \Bigr)+{\rm h.c.} |\Psisig\rangle.$$
Exploiting the same rotational invariance as before, fixing $\vsig=(\sqrt{N},0,0,\ldots)$, we must have all $\mu_i,\nu_i=1$.
So, this is a sum of squares of $N$ different entries of $G$, corresponding to $N$ possible choices of $\sigma$; since we use the complex ensemble, this has vanishing mean.
So, with high probability this term is $\nbos^{p-1} O(N^{1/2+\eta})$ for any $\eta>0$.

So, with high probability,
\be
\label{Egeq}
E
\geq E_0-\lambda \nbos^{p-1} O(N^{p/2+\eta})-\nbos^{p-1} O(N^{1/2+\eta})
\ee
for any $\eta>0$.

Hence,
\begin{theorem}
\label{detectodd}
Let Assumption \ref{assmp} hold.
If $\lambda^2 (p-1)! {\nbos \choose p-1} N^{p} - \emax=\omega\Bigl(\frac{\sqrt{J} \nbos^{p/2-1} N^{p/2}}{\sqrt{\log(N)}}\Bigr)$,
then with high probability algorithm \ref{specalg} correctly determines whether $\lambda=0$ or $\lambda=\overline \lambda$.
\begin{proof}
This follows from the estimate of $E$ in Eq.~(\ref{Egeq}) which lowers bounds the largest eigenvalue when $\lambda=\overline \lambda$ and from theorem \ref{eigbthm} which upper bounds $\Vert H(G) \Vert$.
Note that for 
$\lambda=O(N^{-p/4})$, the terms $-\lambda \nbos^{p-1} O(N^{p/2+\eta})-\nbos^{p-1} O(N^{1/2+\eta})$ on the right-hand side of
this equation are negligible as they are $o(\Bigl(\frac{\sqrt{J} \nbos^{p/2-1} N^{p/2}}{\sqrt{\log(N)}}\Bigr)$.
\end{proof}
\end{theorem}

We now consider recovery.  We will first give a more general bound on cross-terms that will be useful.
We have $$H(\Tspike)=H(\lambda \vsig^{\otimes p}) + H(G) + {\rm cross \, terms}.$$
Let us first bound the cross-terms.

\begin{lemma}
\label{ctb}
With high probability, the operator norm of the cross-terms is bounded by
$\lambda N^{p/2} O(N^{(p-1)/4}) \nbos^{p-1}$.
\begin{proof}
The cross-terms are equal to $2H(T_{cross})$ where
$T_{cross}$ is a tensor for even $q=2(p-1)$ which has components
\be
\Bigl(T_{cross}\Bigr)_{\mu_1,\ldots,\mu_{(p-1)/2},\nu_1,\ldots,\nu_{(p-1)/2},\mu_{(p-1)/2+1},\ldots,\mu_p,\nu_{(p-1)/2},\ldots,\nu_p}=\lambda \sum_\sigma T_{\mu_1,\ldots,\mu_{p-1},\sigma} \cdot \Bigl(\prod_{i=1}^{p-1} (\vsig)_{\nu_i}\Bigr) \cdot (\vsig)_\sigma.
\ee
Rotating so that $\vsig=(\sqrt{N},0,0,\ldots)$, we have
\be
\Bigl(T_{cross}\Bigr)_{\mu_1,\ldots,\mu_{(p-1)/2},\nu_1,\ldots,\nu_{(p-1)/2},\mu_{(p-1)/2+1},\ldots,\mu_p,\nu_{(p-1)/2},\ldots,\nu_p}=\lambda N^{p/2} \sum_\sigma  T_{\mu_1,\ldots,\mu_{p-1},1} \prod_{i=1}^{p-1} \delta_{\nu_i,1}.
\ee

Clearly, $\Vert H(T_{cross}) \Vert \leq \nbos^{p-1} \Vert M_{cross} \Vert,$
where $M_{cross}$ is an $N^{p-1}$-by-$N^{p-1}$ matrix, whose entries are determined by the entries of $T_{cross}$ in an obvious way so that the first $p-1$ indices of $T_{cross}$ determine a column index in $M$ and the last $p-1$ indices of $T_{cross}$ determine a row index of $M$.  Regard $T_{\mu_1,\ldots,\mu_{p-1},1}$ as being the entries of some tensor of order $p-1$ and
let $M'$ by the $N^{(p-1)/2}$-by-$N^{(p-1)/2}$ matrix whose entries are determined by the entries of this tensor, again in the obvious way so that the first $(p-1)/2$ indices of the tensor determine a column index and the last $(p-1)/2$ indices determine a row index.

Then,
\be
\Vert M_{cross} \Vert = \lambda N^{p/2} \Vert M' \Vert.
\ee

However, since the entries of $M'$ are independent Gaussian random variables, it follows from standard random matrix theory results~\cite{mehta2004random} that with high probability $\Vert M' \Vert = O(N^{(p-1)/4})$.
Remark: the dependence on $\nbos$ is not tight here.
\end{proof}
\end{lemma}

So as in the even case,
we have
\begin{theorem}
\label{weakerconvodd}
Let Assumption \ref{assmp} hold.
Given any vector $\Psi$ such that $\langle \Psi | H(\Tspike) |\Psi\rangle \geq (1+c') \emax$ for any scalar $c'>0$,
then with high probability
the corresponding single particle density matrix
$\rho_1$ obeys
\be
\label{obeysodd}
\frac{\langle \vsig| \rho_1 |\vsig\rangle}{N} \geq (c'-o(1))\frac{\emax}{E_0}.
\ee

In particular, for $\langle \Psi | H(\Tspike) |\Psi\rangle \geq \ecut$ with $E_0 \geq (1+c) \emax$ we have
$c'\geq c/2$
so with high probability
\be
\label{obeys2odd}
\frac{\langle \vsig| \rho_1 |\vsig\rangle}{N} \geq \frac{1}{2} \frac{c}{1+c}.
\ee

Hence, algorithm \ref{specalg} achieves recovery.
\begin{proof}
We use lemma \ref{ctb} to bound $\Vert H(T_{cross}) \Vert$.  This bound is asymptotically negligible compared to
$\emax$.  So,
we have
$\langle \Psi | H(\lambda \vsig^{\otimes p} |\Psi\rangle+ \langle \Psi| H(G) |\Psi\rangle \geq (1+c-o(1)) \emax$ where the
$o(1)$ denotes the $o(1)$ contribution from the cross-terms.

Then, the rest of the proof is the same as in the even case, except for a replacement of $p/2$ by $p-1$.  In detail:
by theorem \ref{eigbthm}, with high probability  $\langle \Psi| H(G) |\Psi\rangle\leq (1+o(1)) \emax$.
Hence, $\langle \Psi | H(\lambda \vsig^{\otimes p} |\Psi\rangle\geq (c-o(1)) \emax$.

Rotate so that $\vsig=(\sqrt{N},0,0,\ldots)$.
Then, the Hamiltonian $H(\lambda \vsig^{\otimes p})$ is diagonal in the computational basis for the full Hilbert space.
Let $P(m)$ be the probability, for state $\Psi$, that exactly $m$ out of $\nbos$ qudits are in state $|1\rangle$.  Then,
$\langle \Psi | H(\lambda \vsig^{\otimes p})|\Psi\rangle=\sum_m \lambda^2 (p-1)! {m \choose p-1} N^p P(m)
\leq \sum_m m (E_0/\nbos) P(m)=E_0
\langle \vsig| \rho_1 | \vsig\rangle/N$.
\end{proof}
\end{theorem}

\section{Spectrum of Random Hamiltonian}
\label{randspectra}
In this section, we will estimate the eigenvalues of $H(G)$.  We consider first the case of even $p$.  Here our proof is very similar to the of Ref.~\cite{wein2019kikuchi}, though the method here also suggests some heuristics that may lead to a tighter bound in the future. Then, we consider the case of odd $p$ by reducing it to the case of even $p$.

\subsection{Even $p$}
We first consider the case of even $p$.  
Let $Z(\tau,p,N,\nbos)$ denote
$\expec[{\rm tr}(\exp\{\tau H(G)\})]$ for a tensor $G$ of rank $p$, with entries of $G$ chosen from the Gaussian distribution, with given $N,\nbos$, for some real scalar $\tau$.  In this subsection, $G$ may be symmetrized or not, and may be chosen from either the real or complex ensemble.

The main result in this section is the following lemma:
\begin{lemma}
\label{trbound}
For each ensemble, real or complex, symmetrized or not,
we have
\be
Z(\tau,p,N,\nbos) \leq \dn \exp(\tau^2 (J/2)\nbos^{p/2} N^{p/2}),
\ee
where 
$J$ is a scalar depends that implicitly on $p,\nbos,N$ and
tends to some function depending only on $p$ for large $\nbos,N$.  More precisely, $J$ 
 is equal to $(p/2)! {\nbos \choose p/2!}/\nbos^{p/2}+o(1)$ for the real ensemble and is twice that for the complex ensemble.
\begin{proof}
We first give a brief derivation of Eq.~(\ref{deriv}) below which is a standard result using quantum field theory techniques.
Note that $\tau H(G)=H(\tau G)$ and $\tau G$ is a tensor with entries chosen from a Gaussian distribution with zero mean and variance $\tau^2$.
Hence for any $\tau'>\tau$ we have
\be
Z(\tau',p,N,\nbos)=\expec_{G,\eta} [{\rm tr}(\exp\{H(\tau G+\eta)\})],
\ee
where $\expec_{G,\eta}[\ldots]$ denotes the expectation value over $G$ and $\eta$, with the tensor $\eta$ having Gaussian entries with zero mean and variance $(\tau')^2-\tau^2$.
Taking the expectation value over $\eta$, for $\tau'^2-\tau^2$ small, we need to keep only the zeroth and second order terms on the right-hand side.
So, we find that
\begin{eqnarray}
\label{deriv}
&&\partial_\tau^2 Z(\tau,p,N,\nbos) \\ \nonumber &=&
\int_0^1 {\rm d}s_2 \int_0^{s_2} {\rm d}s_1 \,
\expec_{G,\eta}[
{\rm tr}\Bigl(\exp\Bigl\{(1-s_2)\tau H(G)\Bigr\}
H(\eta)
\exp\Bigl\{(s_2-s_1)\tau H(G)\Bigr\}
H(\eta)
   \exp\Bigl\{s_1 \tau H(G)\Bigr\}\Bigr) ].
   \end{eqnarray}
   Using cyclic properties of the trace, this can be simplified to
\be
\label{deriv2}
\partial_\tau^2 Z(\tau,p,N,\nbos)=
\frac{1}{2} \int_0^1 {\rm d}s_1\,
\expec_{G,\eta}[
{\rm tr}\Bigl(\exp\Bigl\{(1-s_1) \tau H(G)\Bigr\}
H(\eta)
\exp\Bigl\{s_1 \tau H(G)\Bigr\}
H(\eta)\Bigr)].
\ee

 We now use a general result.  Consider  any Hermitian $H$ (we will use $H=\tau H(G)$) and any operator $O$ (we will use $O=H(\eta)=H(\eta)^\dagger$) and any $s_1\in[0,1]$.  We claim that
 $${\rm tr}\Bigl(\exp(H) O^\dagger O\Bigr)+O\leftrightarrow O^\dagger \geq {\rm tr}\Bigl(\exp((1-s_1)H) O^\dagger \exp(s_1 H) O\Bigr) + O\leftrightarrow O^\dagger,$$ where $O\leftrightarrow O^\dagger$ indicates the previous term with $O,O^\dagger$ interchanged.  Proof of claim: work in an eigenbasis of $H$.  It suffices to consider the case that $O=|b\rangle\langle a|$ where $|a\rangle,|b\rangle$ are eigenvectors of $H$ with eigenvalues $E_a,E_b$.  Then the the right-hand side is equal to $\exp(s_1 E_a+(1-s_1) E_b)+\exp(s_1 E_b+(1-s_1) E_a)=2\exp((E_a+E_b)/2) \cosh((s_1-1/2)(E_b-E_a)$.  The cosh function is maximized on the interval $[0,1]$ at $s_1=0,1$ when the right-hand side becomes equal to the left-hand side.
 So,
\be
\label{deriv3}
\partial_\tau^2 Z(\tau,p,N,\nbos)\leq 
\frac{1}{2} \expec_{G,\eta}[
{\rm tr}\Bigl(\exp(\tau H(G) ) 
H(\eta)^2 \Bigr)].
\ee 

For the real ensemble without symmetrization, we have $$\expec[H(\eta)^2]=
\sum_{\mu_1,\ldots,\mu_{p/2}}\sum_{\nu_1,\ldots,\nu_{p/2}}
\prod_{i=1}^{p/2} \Bigl( a^\dagger_{\mu_i} a_{\nu_i} \Bigr)
\prod_{i=1}^{p/2} \Bigl( a^\dagger_{\nu_i} a_{\mu_i} \Bigr)\Bigr.$$
To leading order in $N$, we may approximate $\sum_\nu a_\nu a^\dagger_\nu=N$ so
on the given Hilbert space, $\expec[H(\eta)^2]$ is a scalar equal to
$(p/2)! {\nbos \choose p/2} N^{p/2} +O(N^{p/2-1})$.
In general, for the complex or real ensemble, symmetrized or not, we find that 
$\expec[H(\eta)^2]$ is a scalar equal to $J (p/2)! {\nbos \choose p/2} N^{p/2}$, where $J$ obeys the claims of the lemma.
To verify that $\expec[H(\eta)^2]$ is a scalar and to compute the scalar to all orders in $N$, commute the annihilation operators $a_\nu$ to the right.  The result is some linear combination of operators with all annihilation operators to the right of creation operators which can be written in the form $\sum_{\mu_1,\ldots,\mu_k} (\prod_{i=1}^k a^\dagger_{\mu_k}) (\prod_{i=1}^k a_{\mu_k})$, and each such operator is equal to $k! {\nbos \choose k}$.

Hence,
from Eq.~(\ref{deriv3}), $\partial_\tau^2 \log(Z(\tau,p,N,\nbos) \leq (J/2) \nbos^{p/2} N^{p/2}$.
\end{proof}
\end{lemma}

Remark: this result is clearly not tight in the regime where random matrix theory is accurate ($\nbos=p/2$).
It is interesting to see what happens there.  The correlation function $\expec_{G,\eta}[{\rm tr}\Bigl(\exp\{(1-s_1) \tau H(G)\}
H(\eta)
\exp\{s_1 \tau H(G)\}
H(\eta)\Bigr)]$ is not independent of $s_1$, but rather decays as a function of $s_1$ for $s_1\leq 1/2$ (of course, it increases again as $s_1$ becomes larger than $1/2$).  Considering the regime $s_1\ll 1/2$, using the square-root singularity at the edge of the Wigner semi-circle we can estimate that it decays as $s_1^{-3/2}$.  This means that the integral of this correlation function over $s_1$ is dominated by its value for small $s_1$ of order $1/\tau$ so that for $\tau$ large compared to the inverse width of the semi-circle (though of course $\tau$ not too large) the integral becomes of order $1/\tau$.  This is stronger than the upper bounds here where we have bounded by the integral by something independent of $\tau$.
We may guess that a tighter analysis will show that a similar effect will happen in the case of $\nbos\gg p/2$; however, an important difference occurs.  If we take an eigenstate of $H(G)$ with some eigenvalue $\lambda_0$, and apply $H(\eta)$ for random $\eta$, this only changes $p/2$  out of the $\nbos$ qudits in the state.  So, one might guess that the resulting state will have expectation value of $H(G)$ that is $\lambda_0(1-p/(2\nbos))$ rather than an (as in the random matrix case) expectation value of $H(G)$ which is zero.  So, we may guess that the correlation function will be non-negligible for $s_1 \lesssim (\nbos/p)\tau^{-1}$.  A heuristic estimate in this fashion suggests that the lemma below for the eigenvalue is tight up to logarithmic factors.

From lemma \ref{trbound}, the following lemma is an immediate corollary:
\begin{lemma}
\label{eigbound}
Let $\lambda_1$ be the largest eigenvalue of $G$.
Let $$\emax=\sqrt{2J \log(N)} \nbos^{p/4+1/2} N^{p/4}.$$
Then,
for any $x$,
\be
\prob[\lambda_1\geq x] \leq \exp\Bigl(-\frac{x-\emax}{\xi}\Bigr),
\ee
with
\be
\xi=\frac{\sqrt{J} \nbos^{p/4-1/2} N^{p/4}}{\sqrt{2\log(N)}}
\ee
So, for any $\dem$ which is $\omega(\xi)$, with high probability $\lambda_1\leq \emax+\dem$.
\begin{proof}
We have ${\rm tr}(\exp\{\tau H(G)\})\geq \exp(\tau \lambda_1)$.
Hence, for any $x$, $\prob[\lambda_1 \geq x] \leq Z(\tau,p,N,\nbos)/\exp(\tau x)$.
Since $\dn\leq N^\nbos$, minimizing over $\tau$, we find that
$$\prob[\lambda_1 \geq x] \leq \exp\Bigl(\nbos \log(N) - \frac{x^2}{2J \nbos^{p/2} N^{p/2}}\Bigr).$$
For $x=\emax$, the right-hand is equal to $1$ and for $x>\emax$ the right-hand side decays exponentially.
\end{proof}
\end{lemma}

\subsection{Odd $p$}
We now consider the case of odd $p$.
Let $Z(\tau,p,N,\nbos)$ denote
$\expec[{\rm tr}(\exp(\tau H(G)))]$ for a tensor $G$ of rank $p$, with entries of $G$ chosen from the Gaussian distribution, with given $N,\nbos$.  In this subsection, $G$ is complex and not symmetrized.
The Hamiltonian $H(G)$ is given by Eq.~(\ref{sqeven}) for even $p$ and by Eq.~(\ref{sqodd}) for odd $p$.

We will reduce the calculation for odd $p$ to the case for even $p$, up to a bounded error, showing the following
\begin{lemma}
\label{Zbodd}
For odd $p$,
for $\nbos^{p-1} \tau N^{1/3}=o(1)$,
\be
\label{Zboundodd}
Z(\sqrt{2N} \tau,2(p-1),N,\nbos)  \leq Z(\tau,p,N,\nbos)\leq Z(\sqrt{2N} \tau,2(p-1),N,\nbos) 
 \exp(o(N^{(p-1)/2})).
 \ee
 All occurrences of $Z(\cdot)$ in the above equation refer to the complex ensemble without symmetrizing.
 
 Remark: the assumptions require that $\tau$ is $o(1)$; however, that is still large enough $\tau$ to be useful later in bounding the spectrum since the largest eigenvalues of $H(G)$ are typically large compared to $1$.
 \begin{proof}
For arbitrary $H(G)$,
the exponential $\exp\{\tau H(G)\}$ can be expanded as a series
$$1+\tau H(G) + \frac{\tau^2}{2} H(G)^2 + \ldots.$$
We bring the expectation value (over $G$) inside the trace, and compute the expectation value of this series.  This expectation value can be computed using Wick's theorem to compute the expectation value of a moment of a Gaussian distribution.

For odd $p$, each term in $H$ is a sum of two terms, one depending on the product of two tensors $G$ and one depending on the product of two tensors $\overline G$, where the overline denotes complex conjugation.
Hence, the $m$-th order term $\frac{\tau^m}{m!} H(G)^m$ is a sum of $2^m$ terms, corresponding to $m$ different choices of $G$ or $\overline G$ in each $H(G)$); we call each such choice a {\it history}.
Each term is a product of creation and annihilation operators depending on a product of $2m$ tensors, some of which are $G$ and some of which are $\overline G$.
This expectation value of such a term is non-vanishing only if there are $m$ tensors $G$ and $m$ tensors $\overline G$.
In that case, the
expectation value is given by summing over ways of pairing
each tensor $G$ with a
distinct tensor $\overline G$.  There are $m!$ such pairings.
Then, for each such pairing, one computes an operator by taking the operator
 $\frac{\tau^m}{m!} H(G)^m$ and replacing,
for every pair, the two tensors
$G_{\mu_1,\ldots,\mu_p} \overline G_{\nu_1,\ldots,\nu_p}$  in that pair
with a product of $\delta$-functions $$\prod_{a=1}^p \delta_{\mu_a,\nu_a}=\expec[G_{\mu_1,\ldots,\mu_p} \overline G_{\nu_1,\ldots,\nu_p}].$$
Summing this operator over pairings gives the expectation value.

Note also that here we have not symmetrized $G$.  If we were to symmetrize $G$, then
$\expec[G_{\mu_1,\ldots,\mu_p} \overline G_{\nu_1,\ldots,\nu_p}]$ is given by $1/p!$ times a sum of $p!$ different products of $\delta$-functions, i.e. $\sum_{\pi} \delta_{\mu_a,\nu_{\pi(a)}}$, where $\pi$ is a permutation.
This would lead to additional terms that we need to compute and would make the analysis more difficult (though in practice may lead to better performance).

For given $m$, given history and given pairing, let us define a {\it cluster} as follows: define a graph with $2m$ vertices, each vertex corresponding to a single tensor, either $G$ or $\overline G$.  Let there be an edge between any two tensors which both appear in the same term in the Hamiltonian.  Let there also be an edge between any two tensors which are in a pair.  Hence, this is a graph of degree $2$ (we allow the possibility of multiple edges connecting two vertices if we pair two tensor which both appear in the same term in the Hamiltonian).  A {\it cluster} is a connected component of this graph.

We refer to a cluster containing four vertices as a {\it minimal} cluster.  
A cluster with six or more vertices is called a {\it non-minimal} cluster.
Note that there are no clusters containing only two terms because each term in $H(G)$ depends on a product of two tensors $G$ or two tensors $\overline G$; if we had not taken the complex ensemble and instead taken $G$ to be real, we would instead have these clusters with two terms.
We discuss the case of the real ensemble further after the proof of the lemma.

The minimal clusters will turn out to give the dominant order contribution in an expansion in $N$.  The non-minimal clusters will be subleading order.  

We have expressed $\frac{\tau^m}{m!} H(G)^m$ as a sum over histories and pairings and so $\expec[\exp(\tau H(G))]$ is a sum over $m$, histories, and pairings.
Each term in the sum over $m$, histories, and pairings for $\expec[\exp(\tau H(G))]$ is an operator.
Hence,
$Z(\tau,p,N,\nbos)$ is also given by a sum over $m$, histories, and pairings as one may take the trace of each term in
$\expec[\exp(\tau H(G))]$.
Note that each $m$, history, and pairing gives a non-negative contribution to $Z(\tau,p,N,\nbos)$, i.e., it has a non-negative trace.

{\it Lower Bound---}
Now we prove the first inequality in Eq.~(\ref{Zboundodd}), namely
that $Z(\sqrt{2N} \tau,2(p-1),N,\nbos)  \leq Z(\tau,p,N,\nbos)$.
To do this, 
we first define a similar summation over $m$, histories, and pairings to compute $Z(\tau,q,N,\nbos)$ for even $q=2(p-1)$ and then we compare the two summations.

For even $q$, each term in $H$ is a sum of two terms, one depending linearly on a tensor $G$ and one depending on tensor $\overline G$.
As before, the $m$-th order term $\frac{\tau^m}{m!} H(G)^m$ is a sum of $2^m$ {\it histories}, with each history corresponding to $m$ different choices of $G$ or $\overline G$ in each $H(G)$.
Each term is a product of creation and annihilation operators depending on a product of $m$ tensors, some of which are $G$ and some of which are $\overline G$; note the difference here from the odd case as now there are only $m$ tensors, rather than $2m$.
This expectation value of such a term is non-vanishing only if there are $m/2$ tensors $G$ and $m/2$ tensors $\overline G$.
In that case, the
expectation value is given as before by summing over pairings and replacing the two tensors in the pair with $\delta$-functions.

We claim that if we consider $Z(\tau,p,N,\nbos)$ for odd $p$ and consider only the sum of pairings for which all clusters are minimal, this gives precisely the sum of terms for $Z(\sqrt{2N} \tau,2(p-1),N,\nbos)$.  Since all terms contribute non-negatively to the trace, this will prove the first inequality.

To show this claim, let $L_{2(p-1)}$ label a given choice of $m$, 
history, and pairing
for 
$Z(\sqrt{2N} \tau,2(p-1),N,\nbos)$ and let
$L_{p}$ label a given choice of $m$, history, and clusters for $Z(\tau,p,N,\nbos)$ for which all clusters are minimal.
There are $2^{m/2}$ different pairings for the given choice of clusters labelled by $L_p$.

We will construct a one-to-one correspondence between $L_p$ and $L_{2(p-1)}$ and show that the sum of the terms labelled by $L_p$ is equal to the term labelled by $L_{2(p-1)}$, i.e., that they give the same operator, and hence have the same trace.
Consider given $L_p$.
Then, $L_{2(p-1)}$ is as follows.
The value of $m$ is the same.  The history is also the same: for a given sequence of choices of $G$ or $\overline G$ for $L_p$, we use the same sequence for $L_{2(p-1)}$.
Note that in $L_p$, each of the $m$ choices of $G$ or $\overline G$ denotes a choice that both tensors in $H(G)$ are equal to $G$ or both are equal to $\overline G$, while in $L_{2(p-1)}$ each of the $m$ choices is only a choice about a single tensor in $H(G)$.

The pairing is as follows.  We introduce notation to label the tensors in $H(G)^m$.  
For even $p$, we label the tensors by an integer $i\in \{1,2,\ldots,m\}$ depending which of the $m$ factors of $H(G)$ it is in.  
Here, we mean that the tensors appear in the product $H(G)^m=H(G)\cdot H(G)\cdot \ldots \cdot H(G)$; we label each of the different factors in the product $1,2,\ldots,m$ in sequence.
For odd $p$, we label the tensors by a pair $(i,w)$ for $i\in \{1,2,\ldots,m\}$ and $w\in \{1,2\}$.  The integer $i$ labels which of the $m$ factors of $H(G)$ it is in and $w$ labels whether it is the first or second of the two tensors in $H(G)$.
For a pairing for $Z(\tau,p,N,\nbos)$ with all clusters minimal, 
then each cluster is of the form that for some $i,j$ that we pair $(i,1)$ with $(j,1)$ and $(i,2)$ with $(j,2)$ or of the form that we pair $(i,1)$ with $(j,2)$ and $(i,2)$ with $(j,1)$.
For a given choice of clusters,
the corresponding pairing for $L_{2(p-1)}$ pairs $i$ with $j$.

We sketch the claim about the terms.  For a term in $L_p$,
for each cluster we have two tensors $T$ and two tensors $\overline T$.  We replace these tensors with the expectation value
$$\expec[
\sum_{\mu_1,\ldots,\mu_{p-1}} \sum_{\nu_1,\ldots,\nu_{p-1}}  \sum_{\sigma}
T_{\mu_1,\mu_2,\ldots,\mu_{p-1},\sigma} T_{\nu_1,\nu_2,\ldots,\nu_{p-1},\sigma}
\sum_{\alpha_1,\ldots,\alpha_{p-1}} \sum_{\beta_1,\ldots,\beta_{p-1}}  \sum_{\overline \sigma}
\overline T_{\alpha_1,\alpha_2,\ldots,\alpha_{p-1},\overline \sigma} \overline T_{\beta_1,\beta_2,\ldots,\beta_{p-1},\overline \sigma}],$$ which is equal to some product of $\delta$-functions.
This expectation value then multiplies the operators
$\Bigl(\prod_{i=1}^{(p-1)/2} a^{\dagger}_{\mu_i} a^\dagger_{\nu_i} \Bigr)
\Bigl(\prod_{i=(p-1)/2+1}^{p-1}  a_{\mu_i}a_{\nu_i} \Bigr)$ and
$\Bigl(\prod_{i=1}^{(p-1)/2} a^{\dagger}_{\alpha_i} a^\dagger_{\beta_i} \Bigr)
\Bigl(\prod_{i=(p-1)/2+1}^{p-1}  a_{\alpha_i}a_{\beta_i} \Bigr)$ inserted at the appropriate places into the product $H(G)^m$.
The $\delta$-functions constrain $\sigma=\overline \sigma$; summing over this gives a factor of $N$ for each cluster, while there are two pairings for each cluster giving another factor of $2$ for each cluster.  The number of clusters is equal to $m/2$, giving an overall factor $(2N)^{m/2}$.

This proves the first inequality.

{\it Upper bound---}
Now we prove the second inequality in Eq.~(\ref{Zboundodd}).
To do this, we define the following quantity for $q$ which is a multiple of $4$ (note that $q=2(p-1)$ is a multiple of $4$ if $p$ is odd):
\be
Z'(\tau,\tau',q,N,\nbos) \equiv \expec[{\rm tr}(\exp\{\tau H(G)+\tau' H(G')^2\})],
\ee
where the expectation value is over
tensors
$G$ of order $q$ chosen from the complex ensemble and
tensors $G'$ of order $q/2$ chosen also from the complex ensemble (note that $q/2$ is even).
Note that we square $H(G')$ in the exponent.

We will prove that
\be
\label{scndbnd}
Z(\tau,p,N,\nbos) \leq 
Z'(\sqrt{2N} \tau,\tau',2(p-1),N,\nbos)
\ee
for $\tau'=N^{1/3} \tau$.
From this, the second inequality in Eq.~(\ref{Zboundodd}) follows.  To see this,
we have, for any $G'$, $\Vert H(G')^2 \Vert \leq O(\nbos^{p-1}) \Vert G' \Vert^2$  where the operator norm $\Vert H(G')^2 \Vert$ denotes the largest eigenvalue in absolute value and $\Vert G' \Vert$ denotes the largest singular value of $G'$, regarding $G'$ as a matrix of size $N^{q/4}$-by-$N^{q/4}$.
So, by the Golden-Thompson inequality,
\begin{eqnarray}
\expec_G[{\rm tr}(\exp\{\tau H(G)+\tau' H(G')^2\})] &\leq &\expec_G[{\rm tr}(\exp\{\tau H(G)\})] \expec_{G'}[\exp(O(\nbos^{p-1}) \tau' \Vert G' \Vert^2)]
\\ \nonumber
&=& Z(\sqrt{2N} \tau,2(p-1),N,\nbos)  \expec_{G'}[\exp(O(\nbos^{p-1}) \tau' \Vert G' \Vert^2)].
\end{eqnarray}
where the subscript $G$ or $G'$ denotes the expectation over $G$ or $G'$.
The matrix $G'$ has complex entries; we can bound its operator norm by the sum of the operator norms of its Hermitian and anti-Hermitian parts.  
For $\nbos^{p-1}\tau'=o(1)$, we have  $\expec_{G'}[\exp(O(\nbos^{p-1}) \tau' \Vert G' \Vert^2)]=\exp[O(\nbos^{p-1}) \tau' O(N^{q/4})]$
as can be shown using Hermite polynomials~\cite{mehta2004random} to compute the probability distribution of the eigenvalues of a random matrix and using the decay of a Hermite polynomial multiplying a Gaussian (the study of the largest eigenvalue of a random matrix is a rich field if we consider the probability distribution near the edge of the Wigner semi-circle but here we are satisfied to consider the probability distribution for eigenvalues which are some constant factor $>1$ multiplying the edge of the semi-circle).

To show Eq.~(\ref{scndbnd}),
let $L_p$ label the set of terms for 
$Z(\tau,p,N,\nbos)$
with a given choice of $m$, history, clusters (the clusters need not be minimal), and choice of pairing for all tensors in non-minimal clusters (we do not specify the pairing of the tensors in the minimal clusters; if there are $n_{min}$ minimal clusters then there are $2^{n_{min}}$ terms in the set).
Let $L'_{2(p-1)}$ label a term for
$Z'(\sqrt{2N} \tau,\tau',2(p-1),N,\nbos)$ with given $m$, history, clusters, and pairing.
Here, for $Z'$, a {\it history} corresponds to $m$ choices of either $H(G)$ or $H'(G)^2$ and also to choices of $G$ or $\overline G$, or $G'$ or $\overline G'$, for each $H(G)$ or $H(G')$.
We define a map from choice of $L_p$ to choice of $L'_{2(p-1)}$ so that
the sum of terms labelled by $L_p$ is equal to the term labelled by the corresponding $L'_{2(p-1)}$.
This map will be one-to-one but it will not be onto.  
However, since all terms are non-negative, this will establish the needed inequality.

The map is as follows.  The $m$ will be the same. 
Using the notation above, if tensor $(i,1)$ is in a minimal cluster (and hence, so is $(i,2)$) in the set labelled by $L_p$, then in the history for $L'_{2(p-1)}$
 for the $i$-th term
we choose $\tau H(G)$, while if $(i,1)$ is not in a minimal cluster, then we choose $\tau' H(G')^2$.
Given a history labelled by $L_p$, corresponding to a choice of  tensor or its complex conjugate (i.e., either $G$ or $\overline G$) in each of the $m$ different $H(G)$ in a term in the series for $Z(\tau,p,N,\nbos)$, 
 then we make the same choice of tensor or its complex conjugate in each of the $m$ different  $H(G)$ or $H(G')^2$
 in the history labelled by $L'_{2(p-1)}$.  That is, if we choose $G$ in the $i$-th $H(G)$ in some terms in the series for $Z(\tau,p,N,\nbos)$, then we choose $G$ in $H(G)$ or $G'$ in {\it both} terms in $H(G')^2$ and similarly if we choose $\overline G$ we choose $\overline G$ in $H(G)$ and $\overline G'$ in both terms in $H(G')^2$.
 
 We label the tensors in the expansion for $Z'(\sqrt{2N} \tau,\tau',2(p-1),N,\nbos)$ either by an integer $i$, if it appears in $H(G)$, or by a pair $(i,w)$, if it appears in $H(G')^2$, in which case the index $w\in \{1,2\}$ labels which of the two $H(G')$ it is in.
Finally, we define the pairing labelled by $L'_{2(p-1)}$.  
For a minimal cluster labelled by $L_p$, pairing $(j,1)$ and $(i,2)$ with $(j,2)$ or $(i,1)$ with $(j,2)$ and $(i,2)$ with $(j,1)$, then in the pairing labelled by $L'_{2(p-1)}$ we pair $i$ with $j$..

If a cluster is non-minimal, then we simply use the same pairing for 
the corresponding tensors in $L'_{2(p-1)}$.  That is, suppose a cluster pairs $(i_1,w_1)$ with $(i_2,w'_2)$, and pairs $(i_2,w_2)$ with $(i_3,w'_3)$, and so on, where $w'_a=1$ if $w_a=2$ and $w'_a=2$ if $w_a=1$.
Then, we also pair $(i_1,w_1)$ with $(i_2,w'_2)$, and pairs $(i_2,w_2)$ with $(i_3,w'_3)$, and so on.

The smallest non-minimal cluster has six vertices.  In every cluster, minimal or not, there is a sum over some index (for example, the sum over the index $\sigma=\sigma'$ in the lower bound calculation above) which gives a factor $N$.
Thus, taking $\tau'=\tau N^{1/3}$ accounts for this factor.
No factor of $2$ occurs for the non-minimal clusters.
\end{proof}
\end{lemma}

Remark: since we have chosen the entries of $G$ from the complex ensemble, we have that $\expec[H(G)]=0$ for $p$ odd.  If instead we had chosen the entries of $G$ from the real ensemble we would have (considering the specific case $p=3$ for simplicity and not symmetrizing $G$, again for simplicity) a non-vanishing expectation value since
\be
\expec[T_{\mu_1,\mu_2,\sigma} T_{\nu_1,\nu_2,\sigma}]=N \delta_{\mu_1,\nu_1} \delta_{\mu_2,\nu_2},
\ee
so that
$\expec[H(G)] \propto N \sum_{\mu_1,\mu_2} a^\dagger_{\mu_2} a^\dagger_{\mu_2} a_{\mu_1} a_{\mu_1}$.
Such a term is sometimes called a pairing term or a ``cooperon" in the study of disordered system in physics.
In the case $\nbos=2$ (the smallest possible for $p=3$), this term has operator norm $N^2$. This is much larger than $N^{1/2}$ times the expected operator norm of $H(G)$ for $p=2(p-1)=4$, i.e., that expected operator norm is proportional to $N$ by random matrix theory, and $N^2\gg N^{3/2}$.

There may be other ways to deal with this non-vanishing expectation value other than using the complex ensemble.  One way is to use the real ensemble, but to consider the Hamiltonian $H(\Tspike)-M$, where we define $M=N \sum_{\mu_1,\mu_2} a^\dagger_{\mu_2} a^\dagger_{\mu_2} a_{\mu_1} a_{\mu_1}$ for $p=3$.  In this case, the added term $-M$ cancels the expectation value of $H(G)$ term-by-term in the perturbation expansion.  However, if we do this we still have some additional terms when we consider clusters of four or more vertices.
We expect that the clusters of six or more vertices are still negligible, but the structure of the clusters of four vertices becomes more complicated.  We leave the analysis of this case for the future, but we expect that it would work and may be practically useful.

From lemma \ref{Zbodd}, the following lemma is an immediate corollary:
\begin{lemma}
\label{eigboundodd}
Let $\lambda_1$ be the largest eigenvalue of $G$.
Let $$\emax=2\sqrt{J \log(N)} \nbos^{p/2} N^{p/2},$$
where $J$ is the $J$ of lemma \ref{trbound} for $Z(\tau,2(p-1),N,\nbos)$.
Then,
for any $x$,
\be
\prob[\lambda_1\geq x] \leq \exp\Bigl(-\frac{x-\emax}{\xi}\Bigr),
\ee
with
\be
\xi=\frac{\sqrt{J} \nbos^{p/2-1} N^{p/2}}{\sqrt{\log(N)}}.
\ee

So, for any $\dem$ which is $\omega(\xi)$, with high probability $\lambda_1\leq \emax+\dem$.
\begin{proof}
From lemmas \ref{trbound},\ref{Zbodd}, for $\nbos^{p-1} \tau N^{1/3}=o(1)$, we have
$$Z(\tau,p,N,\nbos)\leq  \exp(\tau^2 J \nbos^{p-1} N^{p}) \exp(o(N^{(p-1)/2})).$$

Let $\tau=\nbos^{1-p/2} N^{-p/2} \sqrt{\log(N)}/\sqrt{J}$.  For $\nbos=o(N^{(p/2-1/3)/(1+p/2)} \log(N)^{1/(2+p)}$, the condition
 $\nbos^{p-1} \tau N^{1/3}=o(1)$ holds.
So, after some algebra, for any $x$, the result follows.
\end{proof}
\end{lemma}

It is worth noting that lemma
\ref{Zbodd}
has the following corollary:
\begin{corollary}
\label{rmtcorr}
For odd $p$,
for $\nbos=p-1$,
with high probability $\lambda_1=O(N^{p/2})$.
\begin{proof}
To prove this we use the existence of tighter bounds for even $q=2(p-1)$ when $\nbos=q/2$.
By Eq.~(\ref{Zboundodd}), 
for $\nbos^p \tau N^{1/3}=o(1)$,
we have
$Z(\tau,p,N,\nbos)\leq Z(\sqrt{2N} \tau,2(p-1),N,\nbos) 
 \exp(o(N^{(p-1)/2}))$.
 Since we are considering fixed $\nbos$, this holds for $\tau N^{1/3}=o(1)$.
We have that $Z(\sqrt{2N} \tau,2(p-1),N,p-1)=\expec_G[\exp(\sqrt{2N} \tau H(G)]$, but the Hamiltonian $H(G)$ is a random
matrix of size $N^{p-1}$-by-$N^{p-1}$ chosen from the so-called Gaussian unitary ensemble~\cite{mehta2004random}.
With high probability, the largest eigenvalue of this matrix is $\Theta(N^{(p-1)/2})$.
Further, we can choose $\tau=\omega(\log(N) N^{-p/2})$ so that 
$Z(\sqrt{2N} \tau,2(p-1),N,p-1)=N \exp(O(\tau N^{p/2}))$; for example, this can be shown by using orthogonal polynomials to bound the probability distribution of the largest eigenvalue as discussed previously.
We have $\prob[\lambda_1>x]\leq Z(\tau,p,N,\nbos) \exp(-\tau x)$.
For $x$ sufficiently large compared to $N^{p/2}$, for $\tau=\omega(\log(N) N^{-p/2})$, the right-hand side is $o(1)$.
\end{proof}
\end{corollary}

\section{Quantum and Classical Algorithms}
\label{QCA}
We now discuss the complexity of classical and quantum algorithms to implement the needed linear algebra.  In particular, we need to determine if that largest eigenvalue is larger than $\ecut$ and we need to find some vector in the eigenspace of eigenvalue $\geq \ecut$.  We emphasize that it is not necessary to find the leading eigenvector itself.

We will use $\psitarget$ to denote this leading eigenvector.  Note that if we lower bound the squared overlap of some vector with $\psitarget$, this will lower bound the probability of success of phase estimation.

We have defined algorithm \ref{specalg} so that in the detection step it uses a hard cutoff on eigenvalue: if the leading eigenvalue is $\geq \ecut$ it reports that $\lambda=\overline \lambda$ while otherwise it reports that $\lambda=0$.  However, no algorithm, classical or quantum, will be able to compute the leading eigenvalue exactly; there will always be some limit on the precision.  Fortunately, the proofs of theorems \ref{detecteven},\ref{detectodd} show that if $E_0\geq (1+c)\emax$ for any $c>0$, then if $\lambda=\overline \lambda$ then with high probability  $\lambda_1\geq
(1-\eta) E_0 + \eta \emax$ for any $0<\eta<1$ while if $\lambda=0$ then with high probability $\lambda_1\leq (1-\eta') E_0 + \eta' \emax$ for any $0<\eta'<1$.  For example, we might take $\eta=1/8$ and $\eta'=1/2$.  So, it suffices instead to implement some ``soft" estimate of the leading eigenvalue which will be very likely to give one result (i.e., reporting that $\lambda=\overline \lambda$)
 if the leading eigenvalue is larger than $(7/8) E_0 + (1/8) \emax$ but very unlikely to give that result if the leading eigenvalue is $\leq \ecut$.

In the quantum algorithms, to obtain some vector in the eigenspace of eigenvalue $\geq \ecut$ and to do this soft estimate, we will implement an approximate projector by phase estimation in the quantum algorithms onto the eigenspace of eigenvalue $\geq (E_0+\ecut)/2=(3/4) E_0+(1/4) \ecut$.  By doing this, the phase estimation error will become negligible when considering the projection of the resulting vector onto the eigenspace with eigenvalue $<\ecut$.  Similarly, in the classical power method we will take the number of iterations sufficiently large that the vector has negligible projection on the eigenspace with eigenvalue $<\ecut$ and further so that it has expectation value for $H(\Tspike)$ greater than $\ecut$.

We begin with a description of some classical algorithms.
The time and space requirements for the classical algorithms are of course not intended to represent a lower bound; rather, they represent times that can be achieved using standard algorithms in the literature.
We then give quantum algorithms.  Finally, we give a further improvement to the quantum algorithm that may be useful in practice.

When we refer to ``space" in a classical algorithm, if we store a $\dn$-dimensional vector, the space requirement is equal to $\dn$ multiplied by the number of bits to store a single entry of the vector.  
In the classical algorithms, we will not discuss issues with finite precision arithmetic in detail.  Since we will be applying operators of the form $H(\Tspike)^m$ to vectors in the ``path integral" methods, we might need to approximate each entry of the vector to accuracy $\dn^{-1} \Vert H(\Tspike) \Vert^{-m}=O({\rm poly}(N^{\nbos m}))^{-1}$.  However, the required number of bits is then only $O(m \nbos \log(N))$ and $m$ will be logarithmic in $N$ so the required number of bits will be only polylogarithmic in $N$.

\subsection{Classical Algorithms}
Classically, the most obvious algorithm is to perform an eigendecomposition on $H(\Tspike)$.  This requires storing matrices of size $\dn$-by-$\dn$ so that the space required is $\tilde O(\dn^2)$ and the time is $\dn^\omega$ where $\omega$ is the matrix multiplication exponent~\cite{Demmel_2007}, though of course in practice the time is closer to $\dn^3$.

However, there is no need to perform a full eigendecomposition.  One can instead initialize a random vector and then apply the power method to extract some eigenvector of $H(\Tspike)$ in the eigenspace with eigenvalue $\geq \ecut$.
 The space required is then only $\tilde O(\dn)$.  The time required for a single iteration of the power method is $\tilde O(\dn)$.  If $\lambda=\overline \lambda$, then in
 $O(\log(\dn)/\log(E_0/\emax))$ iterations, the resulting vector will have an
 $1-o(1)$ projection onto the eigenspace with eigenvalue $\geq (E_0+\ecut)/2$ and a negligible projection onto the eigenspace with eigenvalue $<\ecut$.  So, after this many iterations, one can compute the expectation value of $H(\Tspike)$ on that vector to perform detection, i.e., if $\lambda=\overline \lambda$ the expectation will be larger than $\ecut$ but if $\lambda=0$ the expectation will be close to $\emax$,
 and one can compute the single particle density matrix of the vector after those iterations.
So, the time is
$\tilde O(\dn)O(1/\log(E_0/\emax))$.

This power method still requires exponential (in $\nbos$) space.  In fact, one can use only polynomial space.
One obvious choice is to perform a ``path integral".  That is, given a random initial state $\psirand$, 
the power method computes the state $H(\Tspike)^m |\psirand\rangle$, for some exponent $m$ which gives the number of iterations in the power method.
So, we wish to compute
$$\frac{\langle \psirand | H(\Tspike)^m \rhoone H(\Tspike)^m| \psirand \rangle}{\langle \psirand | H(\Tspike)^{2m} | \psirand\rangle},$$
where the denominator is to correctly normalize the state. 
Let us choose the initial state to be a (normalized) random state from the basis for the symmetric basis given before.
We make this choice to make the ``path integral" simpler and since this is a complete orthonormal basis of states, a random state from this basis has expected overlap with the largest eigenvector of $H(\Tspike)$ equal to $1/N^{\nbos}$.
Then, both the numerator and denominator above can be expressed by a summation over intermediate states from this basis, requiring only space $\tilde O(\log(\dn) m)$; this summation is a ``path integral".  The time required however is now $\tilde O(\dn^m)$ and so becomes significantly worse if the number of iterations is much more than $1$.

An improvement to this time can be obtained by using the algorithm of theorem 4.1 of Ref.~\cite{aaronson2017complexity}.
This algorithm is expressed in terms of qubits, but we can translate the algorithm to get an algorithm for a single qudit of dimension $\dn$, or, using the tradeoffs discussed there, $\nbos$ qudits each of dimension $N$.  

Let us explain this in detail, following the discussion there.  We wish to compute
$\langle y | C | x \rangle$, for some fixed basis states $x,y$ from the basis above (in our application $x=y$, where $C$ is a ``circuit" of some depth $d$ (for example $C=2m+1$ for the numerator above and $C=m$ for the denominator above).
While in Ref.~\cite{aaronson2017complexity}, a circuit was assumed to be unitary, there in fact is no need to assume that; for non-unitary circuits, the issues with finite precision arithmetic do become more important but as remarked above this can be handled with only polylogarithmic overhead.
For us, a circuit is not necessarily unitary; rather it is built out of a sequence of operations, each of which is either $H(\Tspike)$ or $\rhoone$; more generally a circuit might include any operations built out of some tensor of order $O(1)$ multiplying some number of creation and annihilation operators.

For $d=1$, $\langle y | C| x \rangle$ can be computed computed in time ${\rm poly}(\nbos)$.  For $d>1$,
we have
\begin{eqnarray}
\langle y | C|x \rangle=\sum_z \langle y | C_{[d\leftarrow d/2+1]} | z \rangle \cdot \langle z | C_{[d/2\leftarrow 1]} | x\rangle,
\end{eqnarray}
where the summation is over states $z$ in the basis and where $C_{[d\leftarrow d/2+1]}$ and $C_{[d/2\leftarrow 1]}$ are subcircuits from the second and first half of $C$.  If $F(d)$ is the runtime at depth $d$, we have
$F(d) \leq 2 \cdot \dn \cdot F(\lceil d/2 \rceil)$.  So,
$F(d)\leq {\rm poly}(\nbos) (2\dn)^{\lceil \log(d) \rceil}.$
So, the runtime is
$\tilde O((2\dn)^{\lceil \log(2m+1)\rceil})$.
This is much faster than the path integral method but potentially much slower than the power method, depending on the required $m$, i.e., for $E_0/\emax=\Theta(1)$, we need $m$ proportional to $\log(\dn)$ and so the time required is superpolynomially worse than the time required if one stores the full vector.

\subsection{Quantum Algorithms}
We now discuss quantum algorithms for the same problem.  In contrast to the classical algorithms above, all these algorithms take only polynomial space.
First let us describe a simple algorithm, given as algorithm \ref{qalg}.  We then describe a sequence of improvements.

For algorithm \ref{qalg} and all subsequent algorithms, we analyze assuming that $\lambda=\overline \lambda$, i.e., for purposes of analysis we consider the problem of recovery rather than detection.  All these algorithms report success or failure and, if they fail, they are rerun until they succeed.  We give bounds on the expected runtime under the assumption that $\lambda=\overline \lambda$.  If we consider the problem of detection, and if $\lambda=0$, then the algorithm will not report success in the given runtime (since it will be unable to succeed in a certain phase estimation procedure) and so all these algorithms can also be used for detection by running them for some multiple of the given runtime and reporting that $\lambda=0$ if success is not reported in that time.

\subsubsection{Maximally Entangled (or Maximally Mixed) Input State}
Algorithms \ref{qalg},\ref{qalgamp} in this subsubsection work on a Hilbert space which is the tensor product of two Hilbert spaces, each of which have dimension $\dn$.  (In some versions of the Hamiltonian simulation used in the algorithms, it is convenient to embed the symmetric subspace of dimension $\dn$ within the full Hilbert space.)

We use a tensor product notation $A \otimes B$ to denote an operator that is a tensor product of two operators $A,B$ on the two different tensor factors.

\begin{algorithm}
\caption{Quantum Algorithm (simplest, unamplified version).  This and all other quantum algorithms have the same inputs. outputs, and parameter choices as algorithm \ref{specalg}.}
\begin{itemize}
\item[1.] Prepare a maximally entangled state between the two qudits.

\item[2.] Apply phase estimation using $H(\Tspike) \otimes I$.  Let $\psi$ be the resulting state.
If the resulting eigenvalue is larger than $(E_0+\ecut)/2$, report ``success".
Otherwise, report ``failure".

\item[3.] If success is reported, measure and return $\langle \psi | \rhoone \otimes I | \psi \rangle.$  
\end{itemize}
\label{qalg}
\end{algorithm}

Steps $1-2$ are designed to prepare a state whose density matrix on the first qudit has large projection
onto the eigenspace of eigenvalue $\geq \ecut$.
For purposes of analysis, we trace out the second qudit, so that the input state on the first qudit is a maximally mixed state.
If success is reported then (ignoring phase estimation error) we have indeed projected onto this eigenspace.  Ignoring phase estimation error, the probability of success is $\geq 1/\dn$.

Remark:
we prepare a maximally entangled state between the two qudits so that the density matrix on the first state is maximally mixed; we could equally well modify the algorithm to use only a single qudit (reducing the space by a factor of $2$) and prepare a random state on the first qudit.  
This modification however requires some care when we discuss a version of the algorithm that uses amplitude amplification below.
In the unamplified version, we can choose a new random state (for example, choosing it uniformly from any orthogonal basis) each time we perform phase estimation; however, in the amplified version, we must choose a fixed random initial state on the first qudit and amplify the algorithm with that choice of initial state.
We claim (we omit the proof) that if one picks a tensor product of random single qudit states $|v_1 \otimes v_2 \otimes \ldots \otimes v_{\nbos} \rangle$
where $v_a$ are independently chosen from a Haar uniform distribution on the sphere $|v_a|=1$,
 with high probability the state has squared overlap with the leading eigenvector
close to $N^{-\nbos}$; more precisely, ``close" means that with high probability the logarithm of the squared overlap is at least $-\nbos \log(N)\cdot (1+o(1))$.  Note that we might not have this property of the overlap if we had chosen the initial state from the computational basis, for example.  
Note also that
we can efficiently prepare states from this distribution.
Sketch of proof of claim: consider the logarithm of the probability that the sequence of measurements $|v_i\rangle_i\langle v_i|$ succeeds for $i=1,\ldots,\nbos$ in sequence.  The probability that the $i$-th measurement succeeds, conditioned on the previous measurements succeeding, can be computed from the trace of $|v_i\rangle\langle v_i|$ with the reduced density matrix on the $i$-th qudit of some state (i.e., the leading eigenvector projected by the previous measurements), and for any such reduced density matrix with high probability the logarithm of the trace is at least $-\log(N)\cdot (1+o(1))$.

Step $3$ of the algorithm measures some property of a state in the eigenspace with eigenvalue $\geq \ecut$.  It is possible that each time the algorithm is run, one obtains a different energy measurement and hence a different state, so that measuring this property of the state gives some expectation value of $\rhoone$ in a mixed state.  This does not matter since theorems \ref{weakerconv} or \ref{weakerconvodd} also hold for mixed states.

We explain the measurement in step 3 in more detail below.  The simplest possibility is to simply measure one matrix element of $\rhoone$.  Since there are $N^2$ matrix elements, we then need to repeat the algorithm ${\rm poly}(N,\log(\epsilon))$ times to measure each matrix element to accuracy $\epsilon$.  We explain improvements to this in subsection \ref{fi}.

We explain the phase estimation in more detail below.  First, let us analyze the algorithm in a rough outline.  Let the phase estimation be carried out to a precision sufficiently smaller than $E_0-\emax$. 
To define this, we work in an eigenbasis of $H(\Tspike)$.
Let $\tilde\epsilon$ be a bound on the error probability of the phase estimation.  More precisely, we will say that
phase estimation implements an operator $\epe$ which is diagonal in the eigenbasis such that
on a normalized eigenvector $v_i$ with eigenvalue $\lambda_i$ we have
\begin{eqnarray}
\label{epe}
\lambda_i < \ecut \; \rightarrow \; \langle v_i | \epe | v_i \rangle \leq \tilde \epsilon, \\ \nonumber
\lambda_i > (7/8) E_0 + (1/8) \emax \; \rightarrow \; \langle v_i | \epe | v_i \rangle \geq 1-\tilde \epsilon.
\end{eqnarray}
The reader should appreciate that there are a few energy scales here, chosen rather arbitrarily.  The exact value of the energies are not too important.  We have picked $\ecut$ to be partway between $E_0$ and $\emax$ so that it is very unlikely that the largest eigenvalue of $H(G)$ is above $\ecut$ and also very unlikely that the $\lambda_1<\ecut$.  We have picked the energy cutoff in step 2 to be $(E_0+\ecut)/2$ simply to pick some energy partway between $E_0$ and $\ecut$  so that it is very unlikely that phase estimation reports success on an eigenstate with energy $<\ecut$; see first line of Eq.~(\ref{epe}).  In the last line of Eq.~(\ref{epe}) we wrote  $(7/8) E_0 + (1/8) \emax$ simply to pick some energy scale slightly below $E_0$ above which it is very likely that phase estimation reports success; for us later, it would suffices to have any instead a bound for $\lambda_i \geq (1-\eta) E_0 + \eta \emax$ for any $\eta>0$.  Of course, the two lines in Eq.~(\ref{epe}) are not completely symmetric about $(E_0+\ecut)/2$.

Then, 
choose $\tilde\epsilon=\epsilon/\dn$, so that the algorithm reports success with probability at least $(1-\tilde\epsilon)/\dn$ and, given that the algorithm reports success, the resulting state has projection onto the eigenspace with eigenvalue $\geq \ecut$  which is greater than or equal to
$$\frac{(1-\tilde\epsilon)}{(1-\tilde\epsilon)+(\dn-1)\tilde\epsilon}=1-O(\epsilon),$$
where the first term in the denominator is the probability that it reports success on $\psitarget$ as input and the second term is the probability of reporting success on a state in the eigenspace with eigenvalue $< \ecut$, multiplied by $\dn-1$, i.e., multiplied by an upper bound on the dimensionality of that eigenspace.
Taking $\epsilon\ll 1$, we can obtain a large projection onto the eigenspace with eigenvalue $\geq \ecut$, so that
the cost of phase estimation increases logarithmically with $\dn/\epsilon$.

The success probability for $\epsilon\ll 1$ is greater than or equal to $(1/\dn)(1-\epsilon/\dn)$, so for small $\epsilon$ it is very close to $1/\dn$.
Hence, repeating the algorithm until it succeeds, the expected runtime to obtain a single measurement of one matrix element of $\rhoone$ is bounded the time for phase estimation multiplied by $O(\dn)$.

To perform phase estimation, we use controlled simulation of the Hamiltonian $H(\Tspike)$.  There are a large number of quantum simulation algorithms which would work here, such as Refs.~\cite{QSP,LC16,BerryEtAl2014,TS,BCK15} to name just a few.  There are two broad possibilities.  The first possibility is to work in the symmetric subspace of dimension $\dn$.  In this case, $H(\Tspike)$ is a sparse Hamiltonian, and sparse simulation algorithms apply.  The second possibility is to use the Hamiltonian of Eq.~(\ref{firstquantized}) and embed the symmetric subspace into the full Hilbert space; in this case, $H(\Tspike)$ is a local Hamiltonian, in that each terms acts on a small number of qudits, each of dimension $N$, and local simulation algorithms apply.
The cost for these algorithms to simulate for a time $t$ to error $\tilde \epsilon$ is 
${\rm poly}(t \Vert H(\Tspike) \Vert,\nbos,N,\log(\tilde\epsilon))$.

Using the simplest phase estimation algorithm of Ref.~\cite{kitaev2002classical}, the number of bits that we need to phase estimate is $s=O(\log(\Vert H(\Tspike) \Vert / (E_0-\emax))$.
The most expensive bit to obtain is the least significant bit, since obtaining the $j$-th least significant bit requires simulating 
for a time proportional to $2^{s-j}(E_0-\emax)^{-1}$.
So, we can obtain the least significant bit to error $\tilde\epsilon/2$, then obtain the next least significant bit to error $\tilde\epsilon/4$, and so on, making the total error $\tilde\epsilon$.  Of course, a large number of variations of the Kitaev phase estimation algorithm exist in the literature, and any could be used here.

With high probability, $\Vert H(\Tspike) \Vert$ is ${\rm poly}(N)$.
Thus, with high probability the time for phase estimation is ${\rm poly}(N,\nbos,1/(E_0-\emax),\log(\dn/\epsilon))$, giving an algorithm runtime $$\dn {\rm poly}(N,\nbos,1/(E_0-\emax),\log(\dn/\epsilon)).$$

We can speed this algorithm up quadratically by applying amplitude amplification~\cite{brassard2002quantum}.  Modify the phase estimation step $2$ of algorithm \ref{qalg} so that the algorithm phase estimates the eigenvalue, determines if the eigenvalue is larger than $\ecut$, then uncomputes the eigenvalue, returning just a single bit of success or failure.
See algorithm \ref{qalgamp}.
Then, applying amplitude amplification, with high probability the algorithm succeeds in expected time
$\dn^{1/2} {\rm poly}(N,\nbos,1/(E_0-\emax),\log(\dn/\epsilon)).$
Multiplying by ${\rm poly}(N,1/\epsilon)$ to measure $\rhoone$ to accuracy $\epsilon$, the the expected time is still
$$\dn^{1/2} {\rm poly}(N,\nbos,1/(E_0-\emax),\log(\dn/\epsilon)),$$ 
giving a quadratic time improvement, up to ${\rm poly}(N)$ factors, and an exponential space improvement, over the fastest classical algorithm described above.

\begin{algorithm}
\caption{Quantum Algorithm (amplified version)}
\begin{itemize}
\item[1.] Apply amplitude amplification to steps $1-2$ of algorithm \ref{qalg}, modifying step $2$ to uncompute the eigenvalue and return only success or failure.

\item[2.] If success is reported, measure and return $\langle \psi | \rhoone \otimes I | \psi \rangle.$  
\end{itemize}
\label{qalgamp}
\end{algorithm}

\subsubsection{Chosen Input State: Simple Version}
We can obtain a further quadratic speedup by modifying the initial state that we phase estimate.  
In this subsubsection, let us first explain an algorithm that gives the basic idea of the initial state preparation; we will only be able to prove some slightly weaker results for this algorithm (in particular we will prove a lower bound on the average inverse runtime, rather than an upper bound on the average runtime).  We expect that this is primarily a technical issue and that some concentration of measure argument should allow us to prove an upper bound on the average runtime.  We will then describe in the next subsubsection a modification to the algorithm which avoids this technical difficulty and for which we can prove a quadratic improvement without further assumption.

Instead of a maximally entangled state, we can use the tensor $\Tspike$ to prepare a state with a larger projection onto $\psitarget$. 
In this subsubsection, we will work in the $D^\nbos$-dimensional Hilbert space of Eq.~(\ref{firstquantized}), so we will use $\nbos$ qudits each of dimension $N$.
The unamplified version is algorithm \ref{qalginit} and a version with amplitude amplification is algorithm \ref{qalginitamp}.

The most important new step that must be explained is the initial state preparation (we discuss some other details at the end of this subsubsection).  We use the fact that, given a classical list of amplitudes for some $M$-dimensional vector, with the vector having unit norm, we can prepare a quantum state on an $M$-dimensional qudit with the given amplitudes (up to an ill-defined overall phase, of course) using a quantum circuit of depth $O(M)$ and using $O(M)$ classical computation.  For example, labelling the basis states $|0\rangle, |1\rangle,\ldots, |M-1\rangle$, one can start with initial state $|0\rangle$ and apply a sequence of $M-1$ rotations in the two dimensional subspaces spanned by $i\rangle,|i+1\rangle$ for $i=0,\ldots,M-2$.

\begin{algorithm}
\caption{Quantum Algorithm (improved input state, unamplified version)}
\begin{itemize}
\item[1.] Use $\Tspike$ to prepare the initial state $\psiinput$ of Eq.~(\ref{psiinpdef}).

\item[2.] If the initial state is not in the symmetric subspace, report ``failure".  If the state is in the symmetric subspace,
apply phase estimation using $H(\Tspike)$.  Let $\psi$ be the resulting state.
If the resulting eigenvalue is larger than $(E_0+\ecut)/2$, report ``success".
Otherwise, report ``failure".

\item[3.] If success is reported, measure and return $\langle \psi | \rhoone | \psi \rangle.$  
\end{itemize}
\label{qalginit}
\end{algorithm}

\begin{algorithm}
\caption{Quantum Algorithm (amplified version)}
\begin{itemize}
\item[1.] Apply amplitude amplification to steps $1-2$ of algorithm \ref{qalginit}, modifying step $2$ to uncompute the eigenvalue and uncompute the determination of whether the state is in the symmetric subspace and to return only success or failure.

\item[2.] If success is reported, measure and return $\langle \psi | \rhoone  | \psi \rangle.$  
\end{itemize}
\label{qalginitamp}
\end{algorithm}

In this subsubsection, for simplicity in analysis, we assume that the error $\epsilon$ in phase estimation is small enough to be negligible.

We use the same method to produce the input state for both even and odd $p$.
First consider the case that $\nbos$ is an integer multiple of $p$.
For any tensor $T$ of order $p$, let $|T\rangle$ denote
the vector on $p$ qudits (each of dimension $N$) with amplitudes given by 
the entries of the tensor.  This vector is correctly normalized if $|T|=1$.
We prepare the input state 
\be
\label{psiinpdef}
\psiinput=\frac{1}{|\Tspike|^{\nbos/p}} |\Tspike\rangle^{\otimes \nbos/p}.
\ee
Preparing this state takes circuit depth $O(N^p)$ since we can prepare $\nbos/p$ copies of the state $\frac{1}{|\Tspike|} |\Tspike\rangle$ in parallel.

We want to know the expectation value $\langle \psiinput | \epe | \psiinput \rangle$, but
to get oriented, let us estimate the overlap $\langle \psiinput | \Psisig\rangle$.

We have
\be
\langle \vsig^{\otimes p} | \Tspike \rangle=\lambda N^p +\langle \vsig^{\otimes p} | G \rangle. 
\ee
The probability distribution of
$\langle \vsig^{\otimes p} | G \rangle$ is a Gaussian with zero mean and unit variance, so
with high probability, $\langle \vsig^{\otimes p} | G \rangle$ is $o(\lambda N^p)$; indeed, for any increasing function of $N$ which diverges as $N\rightarrow \infty$, with high probability it is bounded by that function.

Hence, with high probability,
\be
\langle \vsig^{\otimes \nbos} | \Tspike^{\otimes \nbos/p} \rangle=(1-o(1)) \cdot \lambda^{\nbos/p} N^\nbos .
\ee
At the same time,
\be
\langle \Tspike^{\otimes \nbos/p} | \Tspike^{\otimes \nbos/p} \rangle=|\Tspike|^{2\nbos/p}.
\ee
We have
$\expec[|G|^2]=O(N^p),$ where the precise constant in the big-O notation depends on whether we symmetrize $G$ or not and whether we use complex entries or not.  Further, $|G|^2$ is a sum of squares of independent random variables (the entries of $G$).  So, by central limit, 
with high probability $|G|^2$ is bounded by $O(N^p)$.
So, with high probability,
$|\Tspike|^{2\nbos/p}=O(N^{\nbos})$.

So, with high probability,
\be
\label{wouldbe}
\frac{\Bigl| \langle \vsig^{\otimes \nbos} | \Tspike^{\otimes \nbos/p} \rangle \Bigr|^2}
{\langle \vsig^{\otimes \nbos} | \vsig^{\otimes \nbos} \rangle \cdot
\langle \Tspike^{\otimes \nbos/p} | \Tspike^{\otimes \nbos/p} \rangle}\geq
(1-o(1))\cdot \lambda^{2\nbos/p} .
\ee
For $\lambda=C N^{-p/4}$, this is $(1-o(1)) C^{2\nbos/p} N^{-\nbos/2}$.

If $\psitarget$ were equal to $\Psisig=N^{-\nbos/2}|\vsig^{\otimes \nbos}\rangle$, then for fixed $N^{-p/4}/\lambda$, Eq.~(\ref{wouldbe}) would give
a lower bound to the squared overlap of the initial state with $\psitarget$ which would be quadratically better (in terms of its scaling with $N$) than the squared overlap for the maximally entangled input state.  So, after applying amplitude amplification, this would give a quartic improvement over the fastest classical algorithm.

However, $\psitarget$ is not equal to $\Psisig=N^{-\nbos/2}|\vsig^{\otimes \nbos}\rangle$ and so this does not give
a lower bound on $\langle \psitarget | \psiinput  \rangle$.

However, we have:
\begin{lemma}
\label{psisigproj}
Suppose that $E_0\geq (1+c) \emax$.  
Then, the projection of $\Psisig$ onto the eigenspace with eigenvalue $\geq (7/8) E_0+(1/8)\emax$ is greater than or equal to
 $\Omega(1) \cdot c/(1+c)$.
\begin{proof}
The expectation value $\langle \Psisig | H(\Tspike) | \Psisig \rangle$ was estimated in subsections \ref{evenpc},\ref{oddpc}.
We have that with high probability, this expectation value is $\geq (1-o(1)) E_0$.
With high probability, the largest eigenvalue of $H(\Tspike)$ in absolute value is bounded by $(1+o(1))(E_0+\emax)$; for the even case this is just the triangle inequality, while for the odd case this uses lemma \ref{ctb}.
Hence, by Markov's inequality applied to $\lambda_1-H(\Tspike)$, the projection of $\Psisig$ onto the eigenspace with eigenvalue $\geq (7/8) E_0+(1/8)\emax$ is greater than or equal to
$(1-o(1)) E_0-((7/8)E_0+(1/8)\emax)/((1+o(1))(E_0+\emax))$ which is $\geq \Omega(1) \cdot c/(1+c)$.
\end{proof}
\end{lemma}

So, $\Psisig$ has some non-negligible projection onto the desired eigenspace.
This does not however yet give us a lower bound on
$\langle \psiinput | \epe | \psiinput \rangle$: we can expand $\psiinput$ as a linear combination of $\Psisig$ and some orthogonal state but we have not bounded the cross-terms in the expectation value.

However, we now give heuristic evidence (not a proof) for a lower bound on
$\expec[\langle \psiinput | \epe | \psiinput \rangle]$.
The main assumption we will make is that $\lambda_1 = E_0 \cdot (1+o(1/\log(N))$; we expect that that assumption can be proven to hold with high probability.
We consider just the case of even $p$
 (we expect that odd $p$ can be handled similarly).
 
 The main reason that we do not give a full proof is that a lower bound on $\expec[\langle \psiinput | \epe | \psiinput \rangle]$ will only imply a lower bound on the expected inverse squared runtime, rather than an upper bound on the expected runtime, which is what we really want.  Instead in the next subsubsection we give a modified algorithm with an upper bound on the expected runtime.
 Let us note that we conjecture that algorithm \ref{qalginitamp} does have a quartic speedup for the expect runtime with high probability.  One might guess that this could be proven using the bound on expectation value of the inverse squared runtime and some concentration of measure argument.  However, we have not been able to make this precise.

 Let us clarify some terminology to distinguish two meanings of the word ``expected", corresponding to averages over $G$ or averages over outcomes of a quantum algorithm, i.e., to the ``expected runtime".
 From here on, when we refer to the ``runtime" of a phase estimation algorithm, this is a short way of saying the expected runtime for a given choice of $G$.  When we refer to the ``expectation value of the runtime", we mean the expectation value over $G$ of this expected runtime.  
Applying amplitude amplification, the runtime is bounded by the time for the state preparation and phase estimation multiplied by the inverse square-root of $\langle \psiinput | \epe | \psiinput \rangle$.  So, lower bounding $\expec_G[\langle \psiinput | \epe | \psiinput \rangle]$ will upper bound the expectation value over $G$ of the inverse squared runtime and we will find a bound on the expectation value of the inverse squared runtime
by
$$\Bigl(N^{\nbos/4} {\rm poly}(N,\nbos,1/(E_0-\emax),\log(\dn/\epsilon))\Bigr)^{-2} \Bigl(N^{-p/4}/\lambda\Bigr)^{-2\nbos/p}$$ in the case that
$E_0\geq E_{max}\cdot (1+c)$ for any $c>0$.
For fixed $N^{-p/4}/\lambda$, this gives a further quadratic improvement, in terms of the scaling of the runtime with $N$, over algorithm \ref{qalgamp}.

Given the existence of that modified algorithm of subsubsection \ref{cimv},
we will just sketch an outline of a possible proof of the lower bound on $\expec[\langle \psiinput | \epe | \psiinput \rangle]$, leaving several details out.  The basic idea is to lower bound this expectation value by a tensor network, working in some approximation which amounts to ignoring fluctuations in $\lambda_1$, then average the tensor network by a sum of pairings, and use this sum to lower bound the expectation value.  Roughly the physical idea is that the terms in $\psiinput$ proportional to $G$ will tend on average to {\it increase} the overlap $\langle \psiinput | \psitarget \rangle$, rather than decrease it.

Consider the operator  $\lambda_1^{-m} H(\Tspike)^m$ for large $m$.  If we take $m$ sufficiently large compared to $\nbos \log(N)/\log(E_0/\emax)$, this will operator will lower bound $\epe$ up to some negligible error.
That is, we take $m$ large enough that $\lambda_1^{-m} H(\Tspike)^m$ is negligibly small acting on any eigenvector
$v_i$ with eigenvalue $\lambda_i \leq (7/8) E_0 + (1/8) \emax$, and for $\lambda_i \geq (7/8) E_0 + (1/8) \emax$, Eq.~(\ref{epe}) gives a lower bound on $\epe$ that is equal to $1$ up to some negligible phase estimation error $\tilde \epsilon$ while clearly $\lambda_1^{-m} H(\Tspike)^m \leq 1$.
For fixed $E_0/\emax$, it suffices to take
$m=O(\nbos \log(N))$.

 For the range of $m$ that we consider here, by assumption we can ignore the fluctuations in $\lambda_1$, i.e., we approximate 
\begin{eqnarray}
|\langle \psiinput | \epe | \psiinput \rangle|^2 & \gtrsim & (\expec[\lambda_1])^{-m}  \expec[\langle \psiinput | H(\Tspike)^m | \psiinput \rangle] \\
&\approx &  (\expec[\lambda_1])^{-m}  \frac{1}{\expec[\langle \Tspike^{\otimes \nbos/p} | \Tspike^{\otimes \nbos/p} \rangle   ]} [\expec[\langle \Tspike^{\otimes \nbos/p} | H(\Tspike)^m | \Tspike^{\otimes \nbos} \rangle],
\end{eqnarray}
where in the second line we further approximate that we can ignore fluctuations in the norm 
$\langle \Tspike^{\otimes \nbos/p} | \Tspike^{\otimes \nbos/p} \rangle$ and treat it as a constant (the proof is a standard large deviation argument on the norm of the tensor).

The quantity  $\langle \Tspike^{\otimes \nbos/p} | H(\Tspike)^m | \Tspike^{\otimes \nbos/p} \rangle$ can be evaluated by a sum of tensor networks, using the full Hilbert space, i.e., 
for each of $m$ choices of $i_1,\ldots,i_{p/2}$ in each of the $m$ factors of $H(\Tspike)$ we have a tensor network).  We can then write each tensor network as a sum of tensor networks, inserting either $\lambda \vsig^{\otimes n}$ or $G$ for each tensor, and we can average these tensor networks over $G$ using the methods of appendix \ref{CTN} by summing over pairings.  Since every term in this sum over networks and pairings is positive, if we restrict to some subset of terms, we get a lower bound.
Let us restrict to the terms in which for $\psiinput$, we choose $\lambda \vsig^{\otimes p}$ for every tensor.  For this set of terms, the tensor network computes precisely $\expec[ \langle \lambda^{\nbos/p} \vsig^{\otimes \nbos} |H(\Tspike)^m | \lambda^{\nbos/p} \vsig^{\otimes \nbos} \rangle]$.
This in turn is $\geq (E_0\cdot (1-o(1/m))^m |\lambda^{\nbos/p} \vsig^{\otimes \nbos}|^2$, since $\langle \Psisig | H(\Tspike) | \Psisig \rangle\geq E_0\cdot (1-o(1/m))$
for $m=O(\nbos\log(N))$.
So the tensor network is lower bounded by $E_0^m |\lambda^{\nbos/p} \vsig^{\otimes \nbos}|^2$.

So, with these approximations we have lower bounded 
$\expec[\langle \psiinput | \epe | \psiinput \rangle]$.

We make some implementation remarks on the algorithm.
The algorithm as described requires measuring whether we are in the symmetric subspace.  Note that the input state $\psiinput$ need not be in the symmetric subspace.
Such a projection can be done for example by phase estimating a Hamiltonian which is a sum of permutation operators.  One can also omit this projection onto the symmetric subspace since our upper bounds on $H(G)$ holds both in the full Hilbert space and in the symmetric subspace.

We have considered the case that $\nbos$ is an integer multiple of $p$.  If $\nbos=kp+l$ for some integers $k,l$ with $0<l<p$, then one can use $l$ ancilla qudits, and prepare an input state which is equal to
$$\frac{1}{|\Tspike|^{k}} |\Tspike\rangle^{\otimes k},$$
on $kp$ qudits, tensored with a maximally entangled state between the remaining $l$ qudits and the remaining $l$ ancillas.  The idea is that we get the additional quadratic improvement in overlap on $kp$ of the qudits, and the remaining $l$ ancilla only cost $1/{\rm poly}(N)$ overlap since $l=O(1)$.

\subsubsection{Chosen Input State: Modified Version}
\label{cimv}
We now modify the algorithm \ref{qalginit} (and its amplified version) to obtain an algorithm for which we can prove the quadratic improvement over algorithm \ref{qalgamp} without any assumption.
Consider given $\Tspike$.  Let $\Delta$ be a $p$-th order tensor, chosen from the same distribution as $G$.
Consider the tensor
\begin{eqnarray}
\Tspike'&=&\Tspike+x\Delta
\\ \nonumber
&\equiv & \lambda \vsig^{\otimes p}+G'
\end{eqnarray}
for some real scalar $x$, where the tensor $G'\equiv G+x\Delta$ has Gaussian entries with variance of the entries equal to $1+x^2$.  We will assume $x=O(1)$; indeed later we will choose $x=o(1)$.  Let us write $\psiinput(T)\equiv |T|^{-\nbos/p} | T^{\otimes \nbos/p}\rangle$ and $\epe(T)$ to denote the phase estimation operator $\epe$ for Hamiltonian $H(T)$.

We have
\begin{eqnarray}
\label{eqex}
G &=&\frac{1}{1+x^2} (G+x\Delta) + \frac{x}{1+x^2}(xG-\Delta) \\ \nonumber
&=& \frac{1}{1+x^2} G' +\frac{x}{\sqrt{1+x^2}} \frac{(xG-\Delta)}{\sqrt{1+x^2}} \\ \nonumber
&=& \frac{1}{1+x^2} G' +\frac{x}{\sqrt{1+x^2}} \delta,
\end{eqnarray}
where $\delta=(1+x^2)^{-1/2}(xG-\Delta)$.
The two random variables $G'$ and $\delta$ have vanishing covariance, so Eq.~(\ref{eqex}) expresses $G$ as
a scalar multiple of $G'$ plus an additional Gaussian random variable $\delta$ which is independent of $G'$.  The variable $\delta$ also has variance $1$.

Given $G$ and $G'$, let
\be
G(y)=y G' + (1-y) G=(y+\frac{1-y}{1+x^2}) G' + \frac{x (1-y)}{\sqrt{1+x^2}} \delta.
\ee
so that $G(y)$ linearly interpolates between $G$ and $G'$.

The idea behind the algorithm is to take a given $G$ as input, randomly perturb to produce $G'$, and then consider several input states $\psiinput(\lambda \vsig^{\otimes p} + G(y))$ with different choices of $y \in [0,1]$.

Let us first recall a property of normalization.  We have 
$\psiinput=\frac{1}{|\Tspike|^{\nbos/p}} |\Tspike\rangle^{\otimes \nbos/p}.$  
Let us write
$$Z=|\Tspike|^{2\nbos/p},$$
so $\psiinput=Z^{-1/2}  |\Tspike\rangle^{\otimes \nbos/p}.$
As shown before, with high probability $Z^{1/2}=|\Tspike|^{\nbos/p}=O(N^{\nbos/2})$.  
Further, with high probability the fluctuations of $Z^{1/2}$ are $o(1)$ compared to its expectation value.
So, from here on we will treat this normalization factor 
$Z$ as a constant, i.e., of course the normalization depends on $N,\nbos$ but we will ignore its dependence on $G$.  
We emphasize that we are not making any additional assumption here as with high probability the fluctuations are asymptotically negligible; we are simply choosing not to write the normalization explicitly.  (Remark: indeed, all we really need is an upper bound on $|\Tspike|^{\nbos/p}$ that holds with high probability,
since the normalization constant $|\Tspike|^{\nbos/p}$ always appears in the denominator.)
Further, we will, without remarking on it further, treat other normalization factors such as $|\psiinput(\lambda \vsig^{\otimes p} + G(y))|$ as constants, and we will introduce other notation for those constants.
Indeed, because we treat the normalization factors such as $Z$ as constants, we will mostly work with un-normalized states which simplifies some of the calculations.

In an abuse of notation,
let us define $\psiinput(y)=\psiinput(\lambda \vsig^{\otimes p} + G(y))$, and write
$Z(y)=|\lambda \vsig^{\otimes p}+G(y)|^{2\nbos/p}$.
Let us write $\Tspike(y)=\lambda\vsig^{\otimes p}+G(y)$.

Let $\Psi(y)$ denote the un-normalized state $|\Tspike(y)\rangle^{\otimes \nbos/p}$ so that $\psiinput(y)=Z(y)^{-1/2} \Psi(y)$.
Let us expand $\Psi(y)$ as a series in $\delta$ and define $\Psi^0(y)$ to denote the zeroth order term in $\delta$.

As a warmup, let us consider $\langle \Psi^0(y) | \epe(\Tspike') | \Psi^0(y) \rangle$.
We consider the higher order terms in $\delta$ later.

We expand the state $\Psi^0(y)$ as a series in $(y+\frac{1-y}{1+x^2})$.
Doing this means that we express $\langle \Psi^0(y) | \epe(\Tspike') | \Psi^0(y) \rangle$ as a polynomial of degree $2\nbos$ which we write as
$\sum_{i \geq 0} a_i (y+\frac{1-y}{1+x^2})^i.$
The zero-th order term $a_0$ is simply equal to
$\langle (\lambda \vsig^{\otimes p})^{\otimes \nbos/p} | \epe(\Tspike') | (\lambda \vsig^{\otimes p})^{\otimes \nbos/p} \rangle$, which is lower bounded in lemma \ref{psisigproj}.
Hence,
\be
\label{a0bound}
a_0 \geq \lambda^{2\nbos/p} N^{\nbos} \Omega(1) \cdot c/(1+c)
\ee
for $E_0\geq (1+c)\emax$.

Now we use a result about polynomials:
\begin{lemma}
\label{polylemma}
Let $p(z)$ be a polynomial of degree $2\nbos$.
Let $[a,b]$ be an interval with $0\leq a <b$.
Then,
\be
{\rm max}_{z\in [a,b]} |p(z)| \geq \Bigl| \frac{a+b}{b-a} \Bigr|^{-2\nbos} \exp(-O(\nbos)) |p(0)|.
\ee
\begin{proof}
We minimize ${\rm max}_{z\in [a,b]} |p(z)|$ over polynomials of degree $2\nbos$ with given value $p(0)$.  
Applying an affine transformation $z\rightarrow (2/(b-a))(z-(a+b)/2)$, this is equivalent to minimizing 
${\rm max}_{z\in [-1,1]} |p(z)|$ over polynomials of degree $2$ with given value $p(z_0)$ for $z_0=(a+b)/(a-b)$.
We claim that this is minimized by $$\frac{p(z_0)}{T_{2\nbos}(z_0)} T_{2\nbos}(z),$$ where $T_{2\nbos}$ is a Chebyshev polynomial.  Proof of claim: suppose some other polynomial $q(z)$ has a smaller maximum absolute value on $[-1,+1]$ with $q(z_0)=p(z_0)$.
Then the polynomial $p(z)-q(z)$ has a zero at $z_0$ but also has at least $2\nbos$ zeros on the interval $[-1,+1]$; this follows from the intermediate value theorem because $T_{2\nbos}$ has $2\nbos+1$ extreme points on the interval which alternate signs.  This gives a contradiction since $p(z)-q(z)$ is degree at most $2\nbos$.

So, ${\rm max}_{z\in [a,b]} |p(z)| \geq |p(z_0)|/|T_{2\nbos}(z_0)|$.
If $0\not\in [a,b]$ then $|z_0|>1$.
We can bound $T_{2\nbos}(z_0)$ for $|z_0|>1$ by $|z_0|^{2\nbos}$ times the sum of absolute values of coefficients of $T_{2\nbos}$.  This sum of coefficients is bounded by $\exp(O(\nbos))$.
\end{proof}
\end{lemma}

We apply this lemma to lower bound ${\rm max}_{y\in [0,1]} |\sum_{i \geq 0} a_i (y+\frac{1-y}{1+x^2})^i|$.  For  $z=y+\frac{1-y}{1+x^2}$, with
$b=1$ and $a=(1+x^2)^{-1}$, this is ${\rm max}_{z\in[a,b]} |p(z)|$ with $p(z)=\sum_i a_i z^i$.
So, using that $(a+b)/(b-a)\leq (2/x^2)$,
we have
\be
{\rm max}_{y\in [0,1]} |\sum_{i \geq 0} a_i (y+\frac{1-y}{1+x^2})^i| \geq
x^{4\nbos} \exp(-O(\nbos)) |a_0|.
\ee

However, for purposes of an algorithm, we need to consider not just the maximum of this polynomial $p(z)$ over an interval, but also consider whether we can come close to this maximum by sampling it at some small number of discrete points.  Fortunately we have the following lemma:
\begin{lemma}
\label{polylemmarandom}
Let $p(z)$ be a polynomial of degree $2\nbos$.
Let $[a,b]$ be an interval which does not contain $0$.
Then, with probability at least $1/2$, if one selects a random point $z$ in the interval $[a,b]$ from the uniform distribution, we have $|p(z)|  \geq \Bigl| \frac{a+b}{b-a} \Bigr|^{-2\nbos} \exp(-O(\nbos)) |p(0)|$.
\begin{proof}
Apply an affine transformation
$z\rightarrow (2/(b-a))(z-(a+b)/2)$, which maps $a$ to $-1$, $b$ to $+1$ and $0$ to $(a+b)/(a-b)$.
Write $p(z)=\prod_i (z-z_i)$ where $z_i$ are zeros of the polynomial; we multiply by $p(z)$ by a normalization constant so that the highest order coefficient is equal to $1$.
Then $\log(|p(z)|)=\sum_i {\rm Re}(\log(z-z_i))$.
Let $A$ denote the average of this logarithm over the interval $[-1,+1]$; by calculus this is
$$A=\frac{1}{2}\sum_i {\rm Re} \Bigl((z-z_i) \log(z-z_i) - (z-z_i)\Bigr)\Bigl|_{z=-1}^{z=+1}.$$
We claim that
\be
\label{avglog}
\log(p((a+b)/(a-b)))-A\leq 2\nbos \cdot \Bigl(\log((a+b)/(a-b))+O(1)\Bigr).
\ee

To show this let $T_i$ denote a given term in the sum for $A$, i.e. $T_i=(1/2){\rm Re} ((z-z_i) \log(z-z_i) - (z-z_i))\Bigl|_{z=-1}^{z=+1}$.
Let $D_i\equiv {\rm Re}(\log((a+b)/(a-b)-z_i)) - T_i$.  By considering various cases, we will show that $D_i\leq \log((a+b)/(a-b))+O(1)$ which implies Eq.~(\ref{avglog}).  
Proof of claim:
for $|z_i|$ less than or equal to some fixed constant (for example, $|z_i| \leq 10$), $T_i$ is lower bounded by some absolute constant, so
$D_i\leq \log((a+b)/(a-b))+O(1)$.
For $|z_i|$ larger than this fixed constant (for example, $|z_i|>10$), $T_i$ is at least $\log(|z_i|)$ minus some other absolute constant,
so $D_i \leq {\rm Re}(\log((a+b)/(a-b)-z_i)-\log(z_i))+O(1)={\rm Re}(\log((a+b)/(a-b)))+O(1)$.

Further, for each $i$, we see that ${\rm max}_{z\in [-1,+1]} {\rm Re}(\log(z-z_i))$ is upper bounded by $T_i+O(1)$.
Hence, ${\rm max}_{z\in[-1,+1]} \log(|p(z)|)$ is upper bounded by $A + O(\nbos)$.
Hence, with probability at least $1/2$, for $z$ chosen uniformly in $[-1,+1]$, we have that
$\log(|p(z)|) \geq A -O(\nbos)$.

Hence, with probability at least $1/2$, we have that $\log(|p(z)|) \geq A-2\nbos \cdot (\log((a+b)/(a-b))-O(\nbos)$.
\end{proof}
\end{lemma}

So, noting that for $z=y+\frac{1-y}{1+x^2}$, a uniform choice of $y$ on $[0,1]$ is the same as a uniform choice of $z$ on $[1/(1+x^2),1]$, we have
\begin{lemma}
For  $y$ chosen randomly from the uniform distribution on $[0,1]$,
the quantity  $\langle \Psi^0(y) | \epe(\Tspike') | \Psi^0(y) \rangle$
is greater than or equal to
$\langle (\lambda \vsig^{\otimes p})^{\otimes \nbos/p} | \epe(\Tspike') | (\lambda \vsig^{\otimes p})^{\otimes \nbos/p} \rangle
\cdot \exp(-O(\nbos)) x^{4\nbos}$.

Using Eq.~(\ref{a0bound}),
\be
\label{xeq}
\langle \Psi^0(y) | \epe(\Tspike') | \Psi^0(y) \rangle
\geq \lambda^{2\nbos/p} N^{\nbos} \exp(-O(\nbos)) x^{4\nbos}\cdot c/(1+c).
\ee
\end{lemma}

We later will worry about optimizing over $x$.  Increasing $x$ will increase the right-hand side of Eq.~(\ref{xeq}),
but it will also change the normalization constant $Z(y)$ that we must multiply by to correctly normalize the input state and also will worsen the threshold for recovery (since $\emax$ will increase).
For now, let us note for orientation that if we choose, for example, $x=0.01$, then the tensor $\Tspike'$ is chosen from the same distribution as $\Tspike$ up to a multiplication by a factor of $(1+0.01^2)^{1/2}$ which is very close to $1$.  This leads to only a very small reduction in the threshold $\lambda$ at which recovery is possible.  At the same time, the change in normalization $(1+0.01^2)^{\nbos}$ is asymptotically negligible (for fixed $\nbos$) compared to the improvement which is polynomial in $N^\nbos$.

We now consider the additional terms depending on $\delta$, i.e., due to the difference $\Psi(y)-\Psi^0(y)$.  At first glance, we have not gained much by the above trick introducing $T'$ since we still have this Gaussian random variable $\delta$ to consider.  However, we have the advantage now that in
$\langle \Psi(y) | \epe(\Tspike') | \Psi(y)  \rangle$, the operator $\epe(\Tspike')$  does not depend on $\delta$ and the states in the bra and ket depend on $\delta$ only polynomially (treating the overall normalization as a constant).

Let us outline our approach.
The main worry that we have is that this quantity $\langle \Psi(y) | \epe(\Tspike') | \Psi(y)  \rangle$ will have a probability distribution that is peaked near some ``small" value, i.e., much less than the value of the overlap at $\delta=0$, rather than some ``large" value, i.e.,
roughly comparable to the value of the overlap at $\delta=0$.
To deal with this, we will use a trick of adding additional noise to make the expectation value of the overlap large, i.e., roughly comparable to the value of the overlap at $\delta=0$.  Now we still need to worry about whether the probability function might have some large probability of having a small value.  So, we will then appeal we appeal to a theorem of Carbery-Wright on ``anti-concentration" of polynomials of Gaussian random variables, bounding the probability that the probability lies in some small interval.  This theorem gives us a useful bound unless the variance of the polynomial is small; however, in that case we can show that the polynomial is likely to be close to its expectation value.

The trick of adding additional noise is as follows.  We perturb the input state by adding additional Gaussian random noise.
Let
$\Psi(y,x')$ denote the unnormalized state
$$|\lambda \vsig^{\otimes p}+(y+\frac{1-y}{1+x^2}) G' + \frac{x (1-y)}{\sqrt{1+x^2}} (\delta+\delta') \rangle^{\otimes \nbos/p},$$
 where $\delta'$ is an additional tensor chosen from a Gaussian distribution with some standard deviation $x'$.
So, the tensor $\delta+\delta'$ is sampled from some distribution with variance $1+x'^2$.
Let $Z(y,x')=|\Psi(y,x')|^2$.
Let $\psiinput(y,x')=Z(y,x')^{-1/2} \Psi(y,x')$ be the ``perturbed input state".

Then, we consider the expectation value over $\delta'$ of $\langle \Psi(y,x') | \epe(\Tspike') | \Psi(y,x') \rangle$; we consider this expectation value as a series in the variable $1+x'^2$.  Using the same treatment as above of this expectation value as a polynomial, we find that if $x'^2$ is chosen uniformly from the interval $[0,x^2]$ then, with probability at least $1/2$, the quantum expectation value $\langle \Psi(y,x') | \epe(\Tspike') | \Psi(y,x') \rangle$ is at least equal to the quantum expectation value with $\delta'=0$ multiplied by 
$\exp(-O(\nbos)) x^{8\nbos}$.

Hence,
\begin{lemma}
\label{expasgood}
With probability at least $1/4$, for uniform random choices of $y$ and $x'^2$ on $[0,1]$ and $[0,x^2]$, the expectation value over $\delta,\delta'$ of
$\langle \Psi(y,x') | \epe(\Tspike') | \Psi(y,x') \rangle$ is at least
$\langle (\lambda \vsig^{\otimes p})^{\otimes \nbos/p} |\epe(\Tspike') | (\lambda \vsig^{\otimes p})^{\otimes \nbos/p}\rangle \cdot \exp(-O(\nbos)) x^{8\nbos}$.
\end{lemma}

We emphasize that throughout we are treating $G'$ as fixed and considering $\delta,\delta'$ as random.  
Now we use anti-concentration in the Gaussian random variables $\delta,\delta'$.
The following lemma is a consequence of the Carbery-Wright theorem~\cite{Carbery_2001}
\begin{lemma}
Let $p(z_1,\ldots,z_n)$ be a polynomial of degree at most $k$ in $n$ independent normally distributed random variables. 
Then, for any $t\in {\mathbb R}$ and any $\delta>0$ with $|t-\expec[p]|>\delta$,
\begin{eqnarray}
\label{cwcomb}
\prob[|p(z_1,\ldots,z_n)-t| \leq \delta] &\leq& 
\Bigl(\frac{\delta}{| t-\expec[p]| -\delta}\Bigr)^{2/(2k+1)} O(k)^{2k/(2k+1)} \\ \nonumber
&\leq &
\Bigl(\frac{\delta}{| t-\expec[p]| -\delta}\Bigr)^{2/(2k+1)} O(k),
 \end{eqnarray}
where
the big-O notation here hides a universal constant.

As a corollary, choosing $t=0$,
\be
\label{ceq}
\prob[|p|\leq \delta |\expec[p]|]\leq \Bigl(\frac{\delta}{1-\delta}\Bigr)^{2/(2k+1)} O(k)=O(\delta^{2/(2k+1)}) O( k).
\ee
\begin{proof}
Let $\Var(p)$ denote the variance of $p(\cdot)$.
As a trivial bound,
\be
\label{bt}
\prob[|p(z_1,\ldots,z_n)-t| \leq \delta] \leq \Var(p)/\Bigl(|t-\expec[p]| -\delta\Bigr)^2.
\ee
By Carbery-Wright,
\be
\label{bcw}
\prob[|p(z_1,\ldots,z_n)-t| \leq \delta] \leq O(k)\cdot (\delta/\sqrt{\Var(p)})^{1/k}.
\ee
Maximizing the bound from Eqs.~(\ref{bt},\ref{bcw}) over $\Var(p)$, Eq.~(\ref{cwcomb}) follows.
\end{proof}
\end{lemma}

Hence, from Eq.~(\ref{ceq}),
\begin{lemma}
\label{lowbndinput}
Consider a choice of $y,x'$ such that the expectation value over $\delta,\delta'$ of the quantum expectation value
$\langle \Psi(y,x') | \epe(\Tspike') | \Psi(y,x') \rangle$
 is at least
$\langle (\lambda \vsig^{\otimes p})^{\otimes \nbos/p} | \epe(\Tspike') | (\lambda \vsig^{\otimes p})^{\otimes \nbos/p} \rangle \cdot \exp(-O(\nbos)) x^{8\nbos}$.
Then, for random choice of $\delta,\delta'$, the quantum expectation value
$\langle \Psi(y,x') | \epe(\Tspike') | \Psi(y,x') \rangle$ is at least
$$\langle (\lambda \vsig^{\otimes p})^{\otimes \nbos/p} | \epe(\Tspike') | (\lambda \vsig^{\otimes p})^{\otimes \nbos/p} \rangle \cdot \exp(-O(\nbos)) x^{8\nbos} \delta,$$
with probability at least $1-O(\nbos) O(\delta)^{2/(2\nbos+1)}$.
\end{lemma}

Lemma \ref{lowbndinput} considers unnormalized input states.  Dividing by normalization constant $Z(y,x')$ we can get a lower bound on the expectation value for the normalized input state $\psiinput(y,x')$ by (for $E_0\geq (1+c)\emax$):
\begin{eqnarray}
&& \frac{1}{Z(y,x')} \langle (\lambda \vsig^{\otimes p})^{\otimes \nbos/p} | \epe(\Tspike') | (\lambda \vsig^{\otimes p})^{\otimes \nbos/p} \rangle \cdot \exp(-O(\nbos)) x^{8\nbos} \delta
\\ \nonumber
&=& \lambda^{2\nbos/p} \exp(-O(\nbos)) x^{8\nbos} \delta \cdot \Omega(1) \\ \nonumber
&=& \Bigl(\frac{\lambda}{N^{-p/4}}\Bigr)^{2\nbos/p} N^{-\nbos/2} x^{8\nbos} \delta \cdot \Omega(1) \\ \nonumber
&=& \Bigl(\frac{\lambda}{N^{-p/4}}\Bigr)^{2\nbos/p} \dn^{-1/2} x^{8\nbos} \delta \cdot \Omega(1).
\end{eqnarray}

Finally, let us pick $x=1/\log(N)$.
Then, we consider the following algorithm \ref{qalm}.

\begin{algorithm}
\caption{Quantum Algorithm (modified improved input state, unamplified version)}
\begin{itemize}
\item[1.] Choose random $y,x'$. Sample $\Tspike',\delta'$ randomly. Prepare input state 
$\psiinput(y,x')$.

\item[2.] If the initial state is not in the symmetric subspace, report ``failure".  If the state is in the symmetric subspace,
apply phase estimation using $H(\Tspike)$  Let $\psi$ be the resulting state.
If the resulting eigenvalue is larger than $(E_0+\ecut)/2$, report ``success".
Otherwise, report ``failure".

\item[3.] If success is reported, measure and return $\langle \psi | \rhoone  | \psi \rangle.$  
\end{itemize}
\label{qalm}
\end{algorithm}

We apply amplitude amplification to algorithm \ref{qalm}.  We apply amplitude amplification under the assumption that indeed 
$\langle \Psi(y,x') | \epe(\Tspike') | \Psi(y,x') \rangle$ is at least
$\langle (\lambda \vsig^{\otimes p})^{\otimes \nbos/p} | \epe(\Tspike') | (\lambda \vsig^{\otimes p})^{\otimes \nbos/p} \rangle
\cdot \exp(-O(\nbos)) x^{4\nbos} \delta.$
From the theorems above, with high probability in $\delta'$, this happens at least $1/4$ of the time.
If the assumption does not hold, we re-sample $y,x'$ and try again.
Then,
 we find that
for $E_0\geq E_{max}\cdot (1+O(1/\log(N))$,
with probability at least $1-O(\nbos) \delta^{2/(2\nbos+1)}$,
the expected runtime of the algorithm is at most
$${\rm poly}(N,\nbos,1/(E_0-\emax,\log(\dn/\epsilon)) \exp(O(\nbos)) \log(N)^{4\nbos} (\frac{N^{-p/4}}{\lambda})^{\nbos/p}
\dn^{1/4}\delta^{-1/2}.$$

Picking $\delta$ to be a slowly decaying function which is $o(1)$, such as $\delta=1/\log(N)$, we find that
\begin{theorem}
\label{impruntimethm}
Let Assumption \ref{assmp} hold.
For $E_0\geq E_{max}\cdot (1+c)$, for any $c>0$, with high probability,
the expected runtime of the algorithm is at most
$${\rm poly}(N,\nbos,1/(E_0-\emax,\log(\dn/\epsilon)) \exp(O(\nbos)) \log(N)^{4\nbos} (\frac{N^{-p/4}}{\lambda})^{\nbos/p}
\dn^{1/4}.$$
\end{theorem}

Remark: the reader may wonder why we picked $x=1/\log(N)$ rather than a more slowly decaying function.  Indeed, any choice of $x$ which is $o(1)$ would suffice, and this would replace the factor $\log(N)^{4\nbos}$ with some other slowly decaying function.

\subsection{Further Improvements}
\label{fi}
In this section, we sketch one improvement that might be practically useful even if it is not theoretically interesting.  For that reason, we do not give any formal proofs in this section.
As we have explained the algorithm above, once the algorithm succeeds in projecting onto the leading eigenvector, it measures some element of the single particle density matrix.  
It is necessary to repeat this operation, including both preparing the leading eigenvector and measuring, a total of ${\rm poly}(N,1/\log(\epsilon))$ times to measure the single particle density matrix to error $\epsilon$, and, as explained, the algorithm needs to try again to prepare the leading eigenvector each time.
The multiplicative overhead
${\rm poly}(N,1/\log(\epsilon))$ can be turned into an additive overhead using the non-destructive measurement technique introduced in Ref.~\cite{Wecker_2015}.  To implement this, we work in the full Hilbert space (the leading eigenvector preparation can be implemented in the symetric subspace, but after preparing the leading eigenvector we work in the full Hilbert space).
Then, we implement a two-outcome projective measurement on a single qudit (this measurement takes us outside the symmetric subspace though it does leave us
in a subspace which is symmetric under permutation of a subset of $\nbos-1$ of the qudits), such as $|\mu\rangle_1\langle \mu|$ for some $\mu$.
Using the non-destructive measurement techniques of Ref.~\cite{Wecker_2015}, we can alternate this two-outcome projective measurement with a projector onto the leading eigenvector (implemented approximately using phase estimation as above) to recover the leading eigenvector.  See Ref.~\cite{Wecker_2015} for more details.

\section{Networks}
\label{networksection}
In this paper, we have used the language of tensor networks to perform some of the estimates.  In this section, we explore the relation to tensor networks further.  Let us consider the simplest task of detection so that the spectral algorithms need only to estimate the largest eigenvalue of $H(\Tspike)$.  In this case, it suffices to evaluate ${\rm tr}(H(\Tspike)^m)$ to sufficiently large power $m$.
This trace can be written as a sum of tensor networks: the sum is present because $H(\Tspike)$ is a sum of terms.  If we restrict to the minimum $\nbos$ (i.e. $\nbos=p/2$ for even $p$ and $\nbos=p-1$ for odd $p$) then no such sum is present.

Consider a given tensor network $N(T)$ in the sum with $N_T$ tensors in the network.  
We write the network $N(T)$ to indicate that the graph and ordering of vertices is given, but that the tensor corresponding to each vertex will be determined by some tensor $T$ (for some vertices it will be $T$ and for some it will be $\overline T$).
We have $N_T=m$ for even $p$ or $2m$ for odd $p$.  Thus, the network has $N_E=p N_T/2$ edges.
The tensor network evaluates to some scalar that we write $\Val(N(T))$.  To get oriented, suppose that we insert $\lambda \vsig^{\otimes p}$ for every tensor in the network. 
In this case, the scalar $\Val(N(\lambda \vsig^{\otimes p}))$ is equal to $\lambda^{N_T} N^{N_E}$.
Suppose instead that we insert $G$ or $\overline G$ for every tensor in the network.  It is possible that $\expec[\Val(N)]$ is zero and it is possible that it is nonzero.  However, let us compute $\expec[|\Val(N(G))|^2]$.  This expectation value can be evaluated using the method of appendix \ref{CTN}.  Note that $\Val(N(G))|^2$ is equal to the evaluation of some tensor network with $2N_T$ tensors; this network is given by two copies of $N$, with all the tensors complex conjugated in one of the two copies.  We refer to the two copies as the first and second copy.  We evaluate this network with $2N_T$ tensors by summing over pairings.  Consider just the pairing in which we pair each tensor in the first copy with the corresponding tensor in the second copy.
It is easy to see that this pairing contributes $N^{N_E}$ to the expectation value.
Hence, since all pairings contribute positively to the expectation value, we have that
$\expec[|\Val(N)|^2]^{1/2} \geq N^{N_E/2}$.

So, we see that
$\expec[|\Val(N(G))|^2]^{1/2} \gg  \Val(N(\lambda \vsig^{\otimes p}))$ for 
$N^{N_E/2} \gg  \lambda^{N_T} N^{N_E}$, i.e., for $\lambda\ll  N^{-p/4}$.
However, this does not yet imply that 
$\expec[|\Val(N(G))|^2]^{1/2} \gg  \expec[\Val(N(\Tspike))]$,
because, in particular,
$\expec[\Val(N(\Tspike))]$ may be larger than 
$\Val(N(\lambda \vsig^{\otimes p}))$.  However,
for any network $N$ suppose we insert $\lambda \vsig^{\otimes p}$ on some set of tensors $S$ in the network and
insert $G$ for the remaining tensors in the network.  In this case, the expectation value is some number given by summing over pairings of the $N_T-q$ insertions of $G$.  For a given pairing $P$, consider the following pairing $Q$ on the tensor network
which computes $\expec[|\Val(N(G))|^2]$.  For each pair of tensors $T_1,T_2$ in $P$, we pair
the first copy of $T_1$ with the first copy of $T_2$ and we pair the second copy of $T_1$ with the second copy of $T_2$.
On the other hand, if a tensor $T$ is not paired in $P$ (so $T$ is in $S$), we pair the first copy of $T$ with the second copy of $T$.
Using this pairing, we again find, for $\lambda\ll N^{-p/4}$, for the given choices of inserting $\lambda \vsig^{\otimes p}$ and $G$ into tensors in the network for $\expec[\Val(N(\Tspike))]$, that the contribution of pairing $Q$ to
$\expec[|\Val(N(G))|^2]$ is much larger than the square of the contribution of pairing $P$ to $\expec[\Val(N(\Tspike))]$.

So, in this regime, the rms fluctuations in the expectation value are large compared to the expectation value itself, i.e., it seems like detection is not possible.
So, to what extent does this result on tensor networks imply that spectral algorithms cannot perform inference for $\lambda<N^{-p/4}$?  Of course, such inference is possible by taking $\nbos$ large so certainly no general impossibility statement can be made.
One way in which one can evade this result is that the result is for a specific tensor network, while the algorithms compute a sum over tensor networks.  Hence, it is possible that by taking a large number of networks in the sum, the rms fluctuations in the sum over networks are small compared to the expectation value of the sum of tensor networks.
This seems to be what happens for the specific algorithms here by taking $\nbos$ large giving a large number of networks in the sum.  This connection to tensor networks may be worth exploring further.

Indeed, the way that we were led to develop the spectral algorithm in this paper is by thinking about different tensor network contractions that might encode useful information for solving inference problems.  After all, a large class of algorithms can be expressed as evaluating in some way a tensor network, and there are many circumstances in which a quantum computer may have some advantage for evaluating a tensor network (given a large number of caveats, for example acting on some vector encoded as a quantum state by a tensor encoded as a quantum operator may be a non-unitary operation which may require some post-selection to implement).  Then, a natural class of tensor networks (in a sense, the ``most symmetric") are those describing the Hamiltonians considered here.  By considering other tensor networks, it may be possible to apply these methods to other inference problems.

\section{Discussion}
\label{discussionsection}
The main new results of this paper are a sequence of spectral algorithms for odd $p$, and a quantum algorithm to compute the single particle density matrix of some vector in an eigenspace with large eigenvalue.  The quantum algorithm could be applied to other matrices other than $H(\Tspike)$ considered here such as $Y$ from Ref.~\cite{wein2019kikuchi}; it is not certain though if the speedup using an input state chosen based on $\Tspike$ would apply to such other matrices.

In this paper, we have related the spectrum for odd $p$ to that for even $q=2(p-1)$, up to an overall scale factor $\sqrt{N}$, at least in the case of Gaussian distributed entries of $G$.  
This enables us to prove correctness of the algorithm for odd $p$.  Also, for $\nbos=p-1$, the spectrum for $q=2(p-1)$ is given by random matrix theory, removing a logarithmic factor present for odd $p$ in some previous work.  See corollary \ref{rmtcorr}.
We have not considered the case of entries of $\Tspike$ being chosen from a discrete distribution, however, we believe that a similar reduction from odd $p$ to even $q=2(p-1)$ could be proven in that case.

We have used entries of $G$ chosen from a complex Gaussian distribution in the case of odd $p$ to simplify the proof of the bounds on the spectrum for odd $p$.  We conjecture however that a modification discussed above to cancel certain ``pairing terms" would work for real entries of $G$.

\bibliographystyle{unsrturl}
\bibliography{spike-ref}
\appendix
\section{Contracting Tensor Networks}
\label{CTN}
In this appendix, we briefly explain how to estimate the average of a tensor network over tensors with entries chosen from a Gaussian distribution, either real or complex, in both cases with zero mean and unit variance.
This is a fairly standard technique, and Ref.~\cite{Hastings_2016mfmc} is an example that uses this technique heavily and can be used as a reference for some of the ideas.

A tensor network is defined by some graph, with each vertex being labelled by some tensor. 
 The degree of each vertex is equal to the order of the corresponding tensor.
  If the tensors are not symmetric, then for each vertex we also specify some ordering on the edges
  attached to that vertex.
  
A tensor network $L$ with $m$ tensors evaluates to a scalar $\Val(L)$ that we will call the value of the tensor network.
The value is computed as follows: for each way of assigning an index to each edge  (there are $N^{pm/2}$ such choices if there are $m$ tensors in the network each of order $p$), 
for every vertex $v$ we compute some scalar; this scalar is equal to a particular entry of the tensor $T^v$ corresponding to that vertex $v$.  The entry that is chosen is given by the indices on the edges attached to that vertex (if the tensor is not symmetric, one uses the ordering on edges of that vertex to determine the
order of indices in the tensor).  Then, the product of this scalar over all vertices $v$ gives some scalar for each way of assigning an index to each edge.  Finally, the sum of this scalar over all possible assignments to edges
 is $\Val(L)$, so that
\be
\Val(L)=\sum_{\rm assignments}\prod_{{\rm vertex} \, v} (T^{\rm v})_{\rm edges},
\ee
where the sum is over choices of indices, the product is over vertices $v$, and $(T^{v})_{\rm edges}$ denotes the entry of a tensor corresponding to the given vertex $v$ determined by the indices on edges attached to that vertex.

Thus given $m$ tensors of order $p$, the value is an $m$-th order polynomial in the entries.  Equivalently, it is a sum over $mp$ different indices (one for each leg attached to each tensor), with $mp/2$ $\delta$-functions (one for each edge) constraining any pair of legs on two different tensors which are connected by an edge to have the same index.

Consider a tensor network in which vertices are labelled by $G$ or $\overline G$ in the complex ensemble or by $G$ in the real ensemble.  We wish to average the value of this network over Gaussian choices of $G$.  
Let us first consider the case in which the tensors are {\it not} symmetrized.
By Wick's theorem, this is computed by a sum over pairings.  Each pairing is either (if all tensors are chosen from a real Gaussian distribution) a way of partitioning the set of tensors into distinct pairs or (if tensors are chosen from the complex distribution) a way of pairing the set into distinct pairs such that we only pair $G$ with $\overline G$.  If there are $m$ tensors in the network, then in the real case there are $m!!$ pairings and in the complex case there are $(m/2)!$ pairings assuming that there are $m/2$ tensors $G$ and $m/2$ tensors $\overline G$ (if not, the expectation value vanishes).

For each pairing, we compute a scalar as follows.  For every pair $T$ and $T'$, replace the two tensors $T_{\mu_1,\ldots,\mu_p}$ and $T'_{\nu_1,\ldots,\nu_p}$ in the polynomial with a product of $\delta$-functions $\prod_{a=1}^{p} \delta_{\mu_a,\nu_a}$.
After making this replacement, sum over all indices.  The result will be equal to $N^{n_{loop}}$ where $n_{loop}$ is some positive integer.  Summing $N^{n_{loop}}$ over all pairings computes the expectation value of the value of the network.

This can conveniently be represented graphically as follows: given a tensor network, and given any tensors that are paired,
``cut" those two tensors out of the network, leaving $2p$ open edges in the network (the edges that were attached to those tensors).  Then, attach those edges to each other, attaching each of the $p$ edges that were attached to the first tensor to the corresponding one that was attached to the second tensor.
After doing this for every pair, what is left is some collection of $n_{loop}$ closed loops.

If the tensors are symmetrized, for each pair we must sum over $p!$ different ways of attaching the edges that were attached to the first tensor to those that were attached to the second tensor.

One can easily also consider the case in which we insert tensors $\vsig^{\otimes p}$ into the network.  This can be done by applying the rotation to make $\vsig=(\sqrt{N},0,0,\ldots)$.  Then, replace each tensor $\vsig^{\otimes p}$ with $\sqrt{N}$ multiplying a product of $\delta$-functions on its edge, $\prod_{a=1}^p \delta_{\mu_a,1}$.  We can represent this graphically simply by writing those $\delta$-functions as open edges; again one sums $N^{n_{loop}}$ over pairings to compute the expectation value of the value of the network, but any open edge with open ends does not count as a closed loop.

\section{Second-Quantized Formalism}
\label{ccr}
In this section, we briefly review the second-quantized formalism for bosons.  This formalism uses a countably infinite basis of states.  In this appendix, we write these basis states as
$|n_1,n_2,\ldots,n_N\rangle$, where each of the $n_\mu$ for $\mu\in \{1,\ldots,N\}$ is a non-negative integer, called an ``occupation number".  We may think of this as labelling possible states of the tensor product of $N$ different harmonic oscillators, i.e., each $n_\mu$ labels a basis state for the $\mu$-th harmonic oscillator, sometimes called the $\mu$-th ``mode".

 The creation operator $a^\dagger_\mu$ is defined by
$$a^\dagger_\mu |n_1,\ldots,n_N\rangle=\sqrt{n_{\mu}+1} |n_1,\ldots,n_{\mu-1},n_{\mu}+1,n_{\mu+1},\ldots,n_N\rangle,$$
i.e., it increases $n_\mu$ by one, leaving the other $n_\nu$ for $\nu\neq \mu$ unchanged, and it multiplies by a scalar $\sqrt{n_{\mu}+1}$.
The annihilation operator $a_{\mu}$ is defined to be the Hermitian conjugate of the creation operator and one may verify that
$$a^\dagger_{\mu} a_{\mu}  |n_1,\ldots,n_N\rangle = n_{\mu}  |n_1,\ldots,n_N\rangle.$$

The number operator $n\equiv \sum_{\mu} a^\dagger_{\mu} a_\mu$ obeys $n |n_1,\ldots,n_N\rangle=\sum_{\mu} n_{\mu} |n_1,\ldots,n_N\rangle$.
In this paper we restrict to the subspace with $n=\nbos$.  This gives us a finite basis of states; indeed, restricting to any finite value for $n$ will give a finite basis.

Finally, the single-particle density matrix is defined by
$$(\rho_1)_{\mu \nu}=\frac{1}{\nbos} \rhoone.
$$

The remarkable property of the second-quantized notation is how it describes behavior in the symmetric subspace of the full Hilbert space.  Any Hamiltonian such as Eq.~(\ref{firstquantized}) which is symmetric under permutation (that is, permuting basis vectors by replacing $|\mu_1\rangle \otimes \ldots \otimes |\mu_{\nbos}\rangle$ with $|\mu_{\pi(1)}\rangle \otimes \ldots \otimes |\mu_{\pi(\nbos)}\rangle$ where $\pi$ is an arbitrary permutation on $\nbos$ elements), will preserve the symmetric subspace, i.e., the subspace which is invariant under permutation.  An othonormal basis for this symmetric subspace can be given by symmetrizing the basis vectors $|\mu_1\rangle \otimes \ldots \otimes |\mu_{\nbos}\rangle$ for $\mu_1\leq \mu_2\leq \ldots \leq \mu_n$; we impose this (arbitrary) ordering on the $\mu$ to avoid redundancies.  The symmetrization of this basis vector is then equal to (using the notation of this appendix) a basis vector of the form $|n_1,n_2,\ldots,n_N\rangle$ where for each $\mu\in {1,\ldots,N}$, the number $n_{\mu}$ is equal to the number of $a\in {1,\ldots,\nbos}$ such that $\mu_a=\mu$.  This basis is also equal to (up to normalization constant) the basis $a^\dagger_{\mu_1} \ldots a^\dagger_{\mu_{\nbos}} |0\rangle$ used in the text.  Then, the single particle density matrix that we have defined above using creation and annihilation operators is equal to (using the formalism of the full Hilbert space) the reduced density matrix (i.e., the marginal) on the first tensor factor in the full Hilbert space (indeed, by symmetry, it is also equal to the marginal on any of the $\nbos$ tensor factors).  Similarly, a symmetric Hamiltonian such as Eq.~(\ref{firstquantized}) can be expressed compactly using creation and annihilation operators as in Eq.~(\ref{sqeven}).
\end{document}